\documentclass[%
 superscriptaddress,
 reprint,
 showpacs,preprintnumbers,
 nofootinbib,
 amsmath,amssymb,
 aps,
 prd,
 floatfix,
]{revtex4-1}
\usepackage{graphicx}
\usepackage{subcaption}
\usepackage{tikz}
\usepackage{xcolor}
\usepackage{hyperref}
\usepackage{comment}
\hypersetup{
    colorlinks=true,
    linkcolor=blue,
    filecolor=blue,      
    urlcolor=blue,
    pdftitle={Study of INCL and NuWro FSI models},
    }

\usepackage{amssymb,amsmath,amstext,amsthm,amsfonts}
\usepackage[utf8]{inputenc}
\usepackage{float}
\usepackage{url}
\usepackage{multirow}
\usepackage{soul}

\RequirePackage{xspace}

\usepackage{lineno}

\definecolor{LightCyan}{rgb}{0.88,1,1}
\usepackage{xcolor,colortbl}
\usetikzlibrary{calc,patterns,angles,quotes}
\begin{document}

\title{Study of final-state interactions of protons in neutrino-nucleus scattering with INCL and NuWro cascade models}


	\author{A.~Ershova}
	\email[Contact e-mail: ]{anna.ershova@cea.fr}
		\affiliation{IRFU, CEA, Universit\'e Paris-Saclay, Gif-sur-Yvette, France}
		
	\author{S.~Bolognesi}
	\email[Contact e-mail: ]{sara.bolognesi@cea.fr}
	\affiliation{IRFU, CEA, Universit\'e Paris-Saclay, Gif-sur-Yvette, France}
	
	\author{A.~Letourneau}
	\email[Contact e-mail: ]{alain.letourneau@cea.fr}
	\affiliation{IRFU, CEA, Universit\'e Paris-Saclay, Gif-sur-Yvette, France}
	
	\author{J.-C. David}
	\affiliation{IRFU, CEA, Universit\'e Paris-Saclay, Gif-sur-Yvette, France}
	
	\author{S.~Dolan}
	\affiliation{European Organization for Nuclear Research (CERN), 1211 Geneva 23, Switzerland}
	
	 \author{J.~Hirtz}    
	\affiliation{IRFU, CEA, Universit\'e Paris-Saclay, Gif-sur-Yvette, France}
	
	\author{K. Niewczas}
	\affiliation{University of Wrocław, Institute of Theoretical Physics, Plac Maxa Borna 9, 50-204 Wrocław, Poland}
	\affiliation{Department of Physics and Astronomy, Ghent University, Proeftuinstraat 86, B-9000 Gent, Belgium}
	
	\author{J.T. Sobczyk}
	\affiliation{University of Wrocław, Institute of Theoretical Physics, Plac Maxa Borna 9, 50-204 Wrocław, Poland}
	
	\author{A.~Blanchet}
	\affiliation{LPNHE, Sorbonne Universit\'e, CNRS/IN2P3, Paris, France}

    \author{M.~Buizza Avanzini}
	\affiliation{Laboratoire Leprince-Ringuet, CNRS, Ecole polytechnique, Institut Polytechnique de Paris, Palaiseau, France}
	
    \author{J.~Chakrani}
	\affiliation{Laboratoire Leprince-Ringuet, CNRS, Ecole polytechnique, Institut Polytechnique de Paris, Palaiseau, France}

	\author{J.~Cugnon}
	\affiliation{AGO department, University of Liège, all\'ee du 6 août 19, bâtiment B5, B-4000 Liège, Belgium}

	\author{C.~Giganti}
	\affiliation{LPNHE, Sorbonne Universit\'e, CNRS/IN2P3, Paris, France}

	\author{S.~Hassani}
	\affiliation{IRFU, CEA, Universit\'e Paris-Saclay, Gif-sur-Yvette, France}

    \author{C.~Juszczak}
	\affiliation{University of Wrocław, Institute of Theoretical Physics, Plac Maxa Borna 9, 50-204 Wrocław, Poland}

   \author{L.~Munteanu }
	\affiliation{IRFU, CEA, Universit\'e Paris-Saclay, Gif-sur-Yvette, France}
	\author{V.Q.~Nguyen}
	\affiliation{LPNHE, Sorbonne Universit\'e, CNRS/IN2P3, Paris, France}

  \author{D.~Sgalaberna }
	\affiliation{ETH Zurich, Institute for Particle Physics and Astrophysics, Zurich, Switzerland}
	
   \author{S.~Suvorov}
	\affiliation{LPNHE, Sorbonne Universit\'e, CNRS/IN2P3, Paris, France}
	\affiliation{Institute for Nuclear Research of the Russian Academy of Sciences, Moscow, Russia}

\begin{abstract}
\noindent 
The modeling of neutrino-nucleus interactions constitutes a challenging source of systematic uncertainty for the extraction of precise values of neutrino oscillation parameters in long-baseline accelerator neutrino experiments. To improve such modeling and minimize the corresponding uncertainties, a new generation of detectors is being developed, which aim to measure the complete final state of particles resulting from neutrino interactions. In order to fully benefit from the improved detector capabilities, precise simulations of the nuclear effects on the final-state nucleons are needed. This article presents the study of the in-medium propagation of knocked-out protons, i.e., final-state interactions (FSI), comparing the NuWro and INCL cascade models. 
The INCL model is used here for the first time to predict exclusive final states in measured neutrino interaction cross sections. This study of INCL in the framework of neutrino interactions features various novelties, including the production of nuclear clusters (e.g., deuterons, $\alpha$ particles) in the final state. The paper includes a complete characterization of the final state after FSI, comparisons to available measurements of single transverse variables, and an assessment of the observability of nuclear clusters.
\end{abstract}

\maketitle

\section{Introduction}
\label{sec:introduction}
The neutrino oscillation physics enters the precision era: notably, the NOVA~\cite{NOvA:2021nfi} and T2K~\cite{patrick_dunne_2020_3959558} long-baseline experiments are featuring measurements of the neutrino mixing angle $\theta_{23}$ and of the largest mass splitting in the atmospheric sector at few percent precision. Sensitivity studies combining future T2K and NOVA data together with measurements at reactor experiments, show the potential of more than 3$\sigma$ significance for possible charge-parity (CP) violation hints and mass ordering determination~\cite{Cabrera:2020own, PhysRevD.103.112010}. The next-generation experiments DUNE~\cite{DUNE:2020lwj} and HyperKamiokande~\cite{Abe:2018uyc} are aiming to establish the mass ordering and possibly discover charge-parity violation with 5$\sigma$ significance, as well as to measure the value of the CP-parameterizing phase ($\delta_{CP}$) with a precision better than 15 degrees. Such results will be enabled by unprecedented statistics of produced and detected neutrinos, requiring an exceptionally robust and precise control of systematic uncertainties.

The largest and most complex systematic uncertainty in present neutrino long-baseline experiments stems from the modeling of neutrino-nucleus interactions. The so called ``near detectors'', placed near the neutrino source before any standard neutrino oscillation can occur, are designed to characterize the neutrino flux and to measure the neutrino-nucleus cross section in order to tune the interaction models and minimize the corresponding uncertainties.
In order to cope with the increasing needs in precision, a new generation of near detectors is being developed based on the concept of a precise and exclusive reconstruction of all the final-state particles produced in neutrino-nucleus interactions. This concept is at the core of the design of the upgrade of T2K near detector ND280~\cite{T2K:2019bbb}. A new active target~\cite{Blondel:2017orl} will allow 3-dimensional reconstruction capabilities for low momentum tracks. Such a detector also enable the reconstruction of neutron kinetic energy by measuring time-of-flight between the neutrino interaction vertex and the first neutron secondary interaction~\cite{Munteanu:2019llq}.
The sensitivity of this type of detector for the most relevant systematic uncertainties in neutrino-nucleus interactions below 1~GeV has been recently documented in Ref.~\cite{Dolan:2021hbw}. Notably,  the exclusive reconstruction of the hadronic part of the final state of neutrino-nucleus interactions permits a more precise reconstruction of the neutrino energy on an event-by-event basis, yet it poses new challenges on the modeling of such interactions. Neutrino-nucleus interaction models have historically been developed and tuned to describe inclusive processes, thus they characterize the cross section as a function of the outgoing lepton kinematics. A new effort is ongoing to expand their predictivity to the kinematics of the hadrons for an exclusive description of the final state~\cite{Dolan:2019bxf,Gonzalez-Jimenez:2021ohu}. 

Most of the available neutrino-interaction models and Monte Carlo simulations describe the initial nuclear state and the fundamental interaction separately from the re-interactions of the final-state hadrons with the nucleus. 
Typically, final-state interactions (FSI) of resulting hadrons are simulated with a cascade mechanism. It is worth mentioning that many nuclear models, originally developed to describe electron-nucleus scattering, include the effect of an outgoing hadrons distortion in a given nuclear potential while solving the lepton-nucleus interaction ~\cite{Benhar:1994hw,Benhar:2005dj,Ankowski:2007uy, Boffi:1993gs, Meucci:2003uy, Meucci:2012yq, Kim:2003mp, Dickhoff:2004xx, Kelly:2005is, Udias:1993xy, Udias:2001tc, Martinez:2005xe, Gonzalez-Jimenez:2012snp, Pandey:2014tza, Nikolakopoulos:2019qcr, Gonzalez-Jimenez:2019qhq, Gonzalez-Jimenez:2019ejf, Nikolakopoulos:2022qkq}. Extensive feasibility studies of implementing such models into generators are nowadays performed in the community, e.g., in Ref.~\cite{Nikolakopoulos:2022qkq}. While such an improvement is certainly needed to cope with the precision of the next generation of long-baseline experiments, we focus on FSI modeling with currently available tools given the needs of running experiments, notably the analysis of the data from the upgraded ND280. 
This paper presents the study of FSI with two different nuclear models implemented in the IntraNuclear Cascade Liege (INCL) code~\cite{Boudard:2012wc, Mancusi:2014fba, Rodriguez-Sanchez:2017odk, Hirtz2018} and in NuWro~\cite{Golan:2012wx}. Fundamental neutrino-nucleus interactions without pion production are studied, where pions in the final state can only be produced through nucleon FSI. This work can be expanded in the future to antineutrino interactions (focusing on neutrons FSI) and to (anti)neutrino interactions with pion production (focusing on pion FSI). The study presented here and its possible further developments aim at evaluating in a detailed way the possible uncertainties inherent to the FSI simulation in neutrino-nucleus scattering. This study is also a step forward into a more precise and complete implementation of FSI effects in Monte Carlo simulations of neutrino-nucleus interactions.

An important novelty in this paper is a discussion of the production of nuclear clusters (like $\alpha$ particles and deuterons) which is modeled in the cascade mechanism of INCL, for the first time, in neutrino-nucleus interactions. Such predictions may significantly impact  the analysis and interpretation of the experimental data, notably for the identification of low momentum particles and the measurement of energy deposits around the neutrino vertex (also known as vertex activity).

The NuWro and INCL nuclear models are described in Section~\ref{sec:nucmod}. The procedure to simulate fundamental neutrino-nucleus interaction events and, subsequently, simulate the intra-nuclear cascade with different models in INCL or NuWro, is described in Section~\ref{sec:sim}. Section~\ref{sec:analysis} introduces the studied variables and the analysis strategy. Section~\ref{sec:res} presents the results from the simulations of the different nuclear models and their comparison. We study different variables to characterize and quantify the impact of FSI on the kinematics of the leading proton, with particular focus on Single Transverse Variables (STV) introduced in Ref.~\cite{Lu:2015tcr} and on the visible energy deposited around the vertex. Section~\ref{sec:data} reports a reasoned comparison of the studied simulations with available measurements of STV.
Finally the main conclusions are reported in Section~\ref{sec:conclusions}.

\section{Nuclear models}
\label{sec:nucmod}

\subsection{NuWro}
Since 2005, the theoretical group of the University of Wrocław has developed NuWro as a comprehensive Monte Carlo lepton-nucleus event generator~\cite{NuWroREPO}, optimizing it for use in accelerator-based neutrino oscillation experiments, i.e., the few-GeV energy region.
Depending on the energy transferred from the interacting leptonic probe to the hadronic system, NuWro provides quasielastic~\cite{Juszczak:2005wk}, hyperon production~\cite{Thorpe:2020tym}, single-pion production and more inelastic channels (DIS)~\cite{Juszczak:2005zs} 
for scattering off free nucleons. After including complex nuclear targets, additional channels such as two-body processes~\cite{Bonus:2020yrd}, coherent pion production~\cite{Berger:2008xs}, and neutrino scattering off atomic electrons~\cite{Zhuridov:2020hqu} are included.
The framework utilizes various nuclear models to provide predictions for the dynamics of target nucleons (e.g., global or local Fermi gas, spectral functions~\cite{Benhar:1994hw,Ankowski:2007uy}, or a momentum-dependent nuclear potential~\cite{Juszczak:2005wk}). Finally, FSI are simulated by an intranuclear cascade which propagates the outgoing nucleons~\cite{Niewczas:2019fro} and produced pions~\cite{Golan:2012wx} through the residual nucleus. 
In the context of this work, the technical aspects of modeling quasielastic neutrino-nucleus scattering and the subsequent FSI are emphasized. More details on the used NuWro version (19.02.2) could be found in Refs.~\cite{Niewczas:2019fro,Niewczas:2020fev}.

In the quasielastic interaction channel, nuclear modeling enters utilizing the {\it plane-wave impulse approximation} that factorizes the one nucleon knock-out processes into the interaction on a single off-shell nucleon convoluted with a particular {\it hole spectral function}, i.e., the probability of leaving the residual system with specific excitation energy and recoil.
The approach relies on the calculation by {\it O. Benhar et al.}~\cite{Benhar:1994hw} that takes into account the electron scattering input to the single-particle wave functions and adding the correlated part evaluated within the local-density approximation.
Additionally, in this model, the prescription by {\it A. Ankowski et al.}~\cite{Ankowski:2014yfa} is applied to go beyond the factorized picture and account for the effects of distorting the final nucleon wave function by an optical potential.
Alternatively, the target nucleons can be treated as constituing the ideal Fermi gas, parametrized through nuclear density or its average value and referred to as {\it local} (LFG) or {\it relativistic Fermi gas} (RFG), respectively.
Finally, the primary interaction vertex is constrained by the conserved vector current (CVC) and partially conserved axial current (PCAC) hypotheses.
The vector form factors are provided by the BBBA05 parametrization~\cite{Bradford:2006yz}, while the axial form factor has a dipole shape with $g_A=1.267$ and the axial mass parameter $M_A = 1.03 \ \mathrm{GeV/c^2}$, according to the discussion in Ref.~\cite{Megias:2019qdv}.

Modeling final-state interactions is a challenging many-body problem that bears a tension between numerical efficiency and accuracy of nuclear calculations.
NuWro solution is based on seminal papers by {\it N. Metropolis et al.}~\cite{Metropolis:1958wvo,Metropolis:1958sb}, which describe an algorithm of the space-like cascade model, and applied up-to-date physics ingredients.
In this approach, mean free paths are attributed to the particles propagated in straight lines with steps of $\Delta x$ through a continuous medium.
Such Monte Carlo sampling uses the standard non-interaction probability formula
\begin{equation}
P(\Delta x) = \exp(-\Delta x/\lambda),    
\end{equation}
where $\lambda = (\rho \sigma)^{-1}$ is the mean free path calculated locally, expressed in nuclear density $\rho$ and an effective interaction cross section $\sigma$.
The maximal step of $\Delta x = 0.2 \ \mathrm{fm}$ is sufficient to grasp the structure of commonly used density profiles.
By default, the nucleons constituting the nuclear medium originate from the LFG model, and therefore, meet its Pauli blocking rules (applied on an event-by-event basis).
The cascade terminates when all the moving hadrons leave the nucleus or do not have enough kinetic energy and are stuck in nuclear potential (with the separation energy of $7 \ \mathrm{MeV}$).
The remnant nucleus is in an excited state, and its deexcitation is not modeled.

In Ref.~\cite{Niewczas:2019fro}, the nucleon part of the NuWro cascade has been exhaustively tested, aiming to reproduce the nuclear transparency in exclusive $(e,e'p)$ scattering experiments.
The essence of this model lies in nucleon-nucleon cross sections, which replicate the PDG dataset~\cite{ParticleDataGroup:2016lqr}, the fraction of single-pion production adjusted to follow the fits of Ref.~\cite{Bystricky}, and the center-of-momentum frame angular distributions of Ref.~\cite{Cugnon:1996kh}.
Additionally, the cross sections are modified with in-medium corrections~\cite{Pandharipande:1992zz,Klakow:1993dj} and two-nucleon correlation effects~\cite{Niewczas:2019fro}.
Finally, the pion-nucleon interaction dynamics is taken from the model of {\it L.L. Salcedo et al.}~\cite{Salcedo:1987md}.
This aspect, together with the formation zone effect for the inelastic scattering channels, has been presented and compared to data in Ref.~\cite{Golan:2012wx}.

\subsection{INCL}

\begin{figure}[ht]
\centering 
\includegraphics[width=0.98\linewidth]{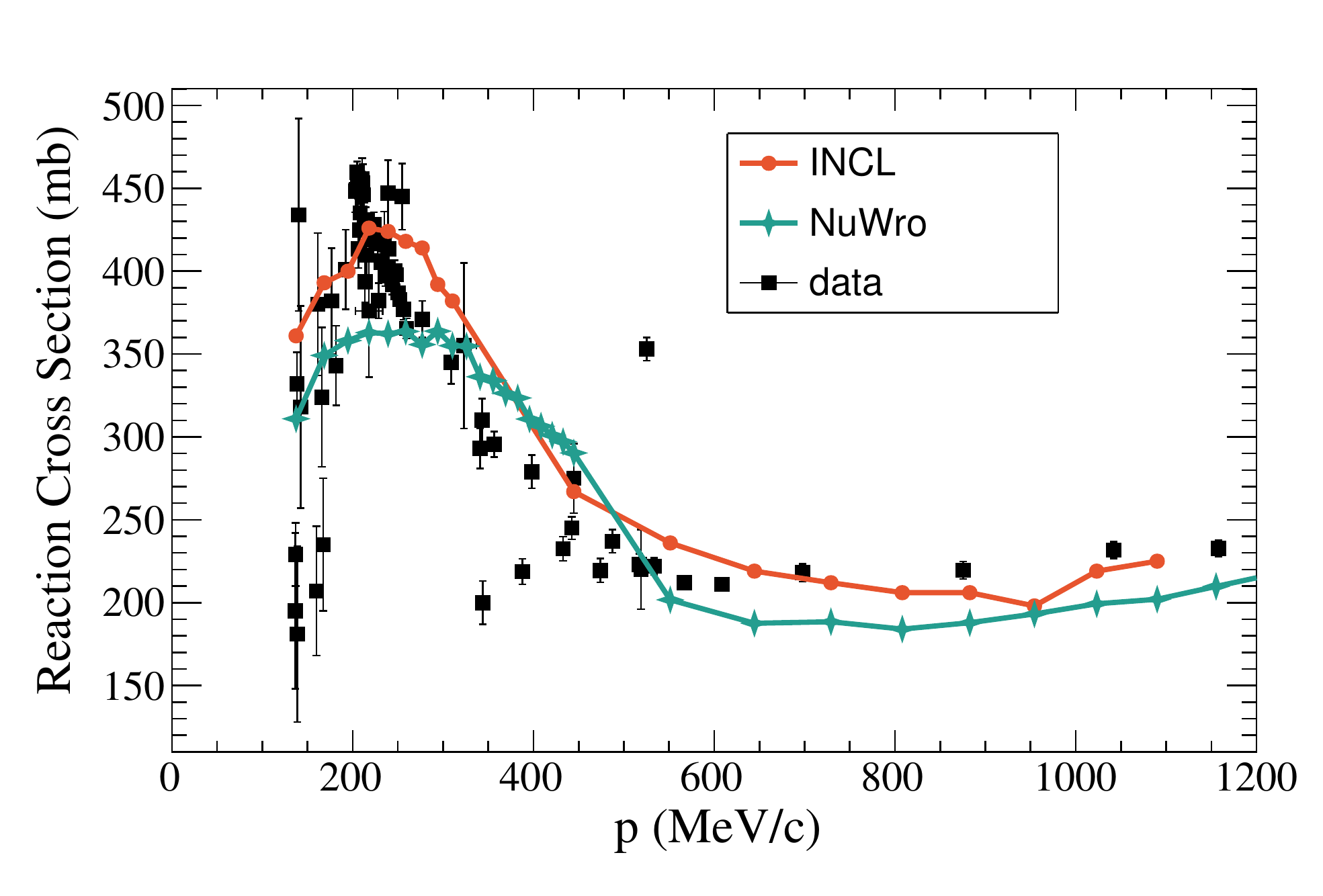}
\caption{\label{fig:p_benchmark} Cross section of proton-$^{12}$C interactions as a function of the proton momentum: comparison of the INCL and NuWro models to available experimental data~\cite{Auce, Ingemarsson:1999sra, Trzaska:1991ww, Slaus:1975zz, Menet1, Menet2, Pollock, Wilkins, Meyer, GOODING1959241, BURGE1959511, Cassels, RENBERG197281, Dicello, Kirkby, MAKINO1964145, TAYLOR1961642, GOLOSKIE1962474, Jaros}.  Results of the NuWro 19.02.2  simulation were taken from Ref.~\cite{PhysRevD.104.053006}.}
\end{figure}

INCL is a nuclear model dedicated to the simulation of the reactions induced by baryons (nucleons, $\Lambda$, $\Sigma$), mesons (pions and Kaons) or light nuclei (A$\leqslant$18) on a target nucleus. It shows a remarkable agreement with an exhaustive list of experimental data of Refs.~\cite{2015, IAEA}. As an example, the comparison to cross section data of proton interactions with fixed $^{12}$C target is shown in Fig.~\ref{fig:p_benchmark}. The INCL cascade is usually coupled to the deexcitation code ABLA~\cite{ABLA} but can be combined also to the SMM~\cite{Botvina:1994vj, Bondorf:1995ua} or GEMINI++~\cite{Mancusi:2010tg, Charity:1988zz} deexcitation codes. INCL is a very flexible tool that has been implemented in GEANT4~\cite{Allison:2016lfl} and GENIE~\cite{PhysRevD.104.053006}.
In this study the stand-alone version of INCL is studied and the subsequent deexcitation model is disabled.

The standard version for the projectile employs the impact parameter formalism taking into account the Coulomb distortion. 
A `working sphere', where all events take place,  is defined with radius $R_{max}$:

\begin{equation}
    R_{max} = \left\{
    \begin{array}{lll}
        R_0 + 8a & \mbox{for } A\: \textgreater\: 19\\
        5.5 + 0.3\left( A - 6 \right)/12 & \mbox{for } 6\: \leq A\: \leq\: 19\\
        R_0 + 4.5 & \mbox{for } A \geq 2
    \end{array}
\right.
\end{equation}
where $R_0$ and $a$ are radius and diffuseness of the target nucleus density respectively. For carbon, $R_{max}$~=~5.7~fm.
 If the trajectory of the projectile, defined by a randomly chosen asymptotic impact parameter and modified by the Coulomb field of the target, intersects the working sphere, the projectile enters the nucleus and can interact with the nucleons or not (a “transparent” event).

The simulation of neutrino interactions in INCL is not yet available. In this study, the neutrino-nucleon interaction inside the nucleus is enforced by using the neutrino vertex from the NuWro code. Then, the produced particles from the neutrino-nucleon interaction may undergo final-state interactions and initiate a cascade.  The exact matching procedure to interface NuWro and INCL will be described in Sec.~\ref{sec:sim}. The overall normalization is taken from the total cross section calculated by NuWro. 

The INCL nuclear model is essentially classical, with some additional ingredients to mimic quantum effects. Each nucleon in the nucleus has its position and momentum and moves freely in a square potential well. The radius of the potential well depends on the kinetic energy of the nucleon. Nucleon momenta are distributed uniformly in a sphere whose radius is defined by the maximal Fermi momentum. Position and momentum are correlated in INCL. In such a {\it classical} picture, the particle is allowed to move in a sphere with the maximum radius fixed by its momentum, and so its position is sampled in this sphere. This picture has been refined to take into account the quantum properties of the wave functions. Based on a HFB (Hartree-Fock-Bogoliubov) formalism, the correlation is still valid but less strict, i.e. the nucleon has a non-zero probability for going beyond the maximum radius. Further details can be found in Ref.~\cite{Rodriguez-Sanchez:2017odk}.

The cascade reaction can be described as an avalanche of independent binary collisions. Different types of events inside the cascade could occur: collisions, decays and, at the surface, reflections or transmissions. The particles travel along straight lines and the different possible fates are calculated:
\begin{itemize}
\setlength\itemsep{-0.4em}
    \item two particles reach the minimal distance to interact,
    \item a particle hits the border of the potential well then it deflects back or leaves the nucleus,
    \item a particle decays (e.g. a $\Delta$ resonance or $\omega$ meson).
\end{itemize}
The fate with the shortest time is chosen.

Nucleons that do not participate in the cascade are defined as \textit{spectators}. In the beginning, all nucleons are spectators except for the projectile. A spectator could become a participant while interacting with the projectile or another participant.
To prevent spontaneous nucleon emission (nuclear boiling), the spectator nucleons of the target cannot interact between themselves. The projectile feels the nuclear potential of the target. A participant nucleon can become a spectator again if its energy decreases below a threshold (details are in~Ref.~\cite{Boudard:2012wc}). While such an option has been included to  improve the agreement with some sets of data~\cite{Boudard:2012wc}, the modeling of very low energy nucleons is still an open issue for cascade mechanisms. In the scope of this paper, this feature is disabled. 
Participants can be absorbed at the end of the cascade when the stopping time is reached and the nucleus is thermalized through equipartition of energy. In this case, the participant's energy is left in the nucleus, putting it in an excited state.

The cascade lasts until one of these conditions is satisfied:
\begin{itemize}
\setlength\itemsep{-0.4em}
    \item the stopping time determined by the model is reached,
    \item there are no more participants,
    \item the mass number of the target is less than 4,
    \item the projectile leaves without an interaction (transparent event).
\end{itemize}
Concerning a participant nucleon at the surface, it can be reflected or emitted. It leaves the nucleus if its energy is higher than the Fermi energy plus the value of its separation energy that is taken from mass tables based on experimental data~\cite{Tuli}. However, a notable feature of INCL is that an outgoing nucleon (subject to FSI) with some probability could clusterize with other nucleons and leave the nucleus as a nuclear cluster (e.g., $\alpha$ particle).  The emission of nuclear clusters was extensively compared with representative experimental data. One can find more information about cluster emission in INCL in Ref.~\cite{Mancusi:2014fba}.

In addition to quantum effects embedded in the elementary cross sections used and the one previously quoted with the refined position-momentum correlation related to the quantum behavior of the wave function, Pauli-blocking is considered and checked. 
If the collision is blocked, the particle propagates until it reaches the next allowed interaction in the cascade and the products of that reaction are further propagated.
There are two main Pauli blocking models implemented in INCL: the strict one, forbidding the interaction if the projectile momentum is below the Fermi momentum, and the statistical model that considers only nearby nucleons in the phase-space volume and acts upon the calculated occupation probability.
The default option applied in INCL is a compromise between the strict Pauli blocking for the first collision and the statistical one for the subsequent ones. Motivation and details of this procedure can be found in Ref.~\cite{Boudard:2012wc, Henrotte}. In this paper, the first interaction is the neutrino interaction with the nucleus taken from NuWro. So for the first interaction, the Pauli blocking from NuWro is used, then for subsequent proton interactions statistical Pauli blocking is applied.

\section{Simulation procedure}

\label{sec:sim}
The study focuses on the comparison between NuWro and INCL of proton-induced FSI cascades in Charged-Current Quasi-Elastic (CCQE) neutrino interactions on Carbon using the T2K neutrino energy flux at the near detector from Ref.~\cite{T2K:2015sqm}. The initial state and the neutrino-nucleus interaction is simulated with NuWro with the Spectral Function (SF) approach. About 350000 CCQE events have been simulated with NuWro. In each event, the leading proton exiting the interaction of a neutrino on a neutron is then injected in the INCL nuclear model and FSI is simulated with INCL. The results are compared to the FSI cascade simulation done in NuWro. 

In the INCL simulation, the neutron with the closest momentum-vector to the initial neutron simulated by NuWro for the neutrino interaction is selected.  This procedure is independent of the exact values of neutron momenta in NuWro and INCL and is applied event-by-event for the whole sample. The chosen INCL neutron is then substituted with the outgoing particles (muon and proton) preserving their kinematic properties calculated from the neutrino-neutron interaction by NuWro. The Pauli blocking in the neutrino interaction is simulated by NuWro, while the Pauli Blocking along the proton FSI cascade is simulated differently by NuWro and INCL, as explained in Sec.~\ref{sec:nucmod}.

\begin{figure}[ht]
\centering 
\includegraphics[width=0.98\linewidth]{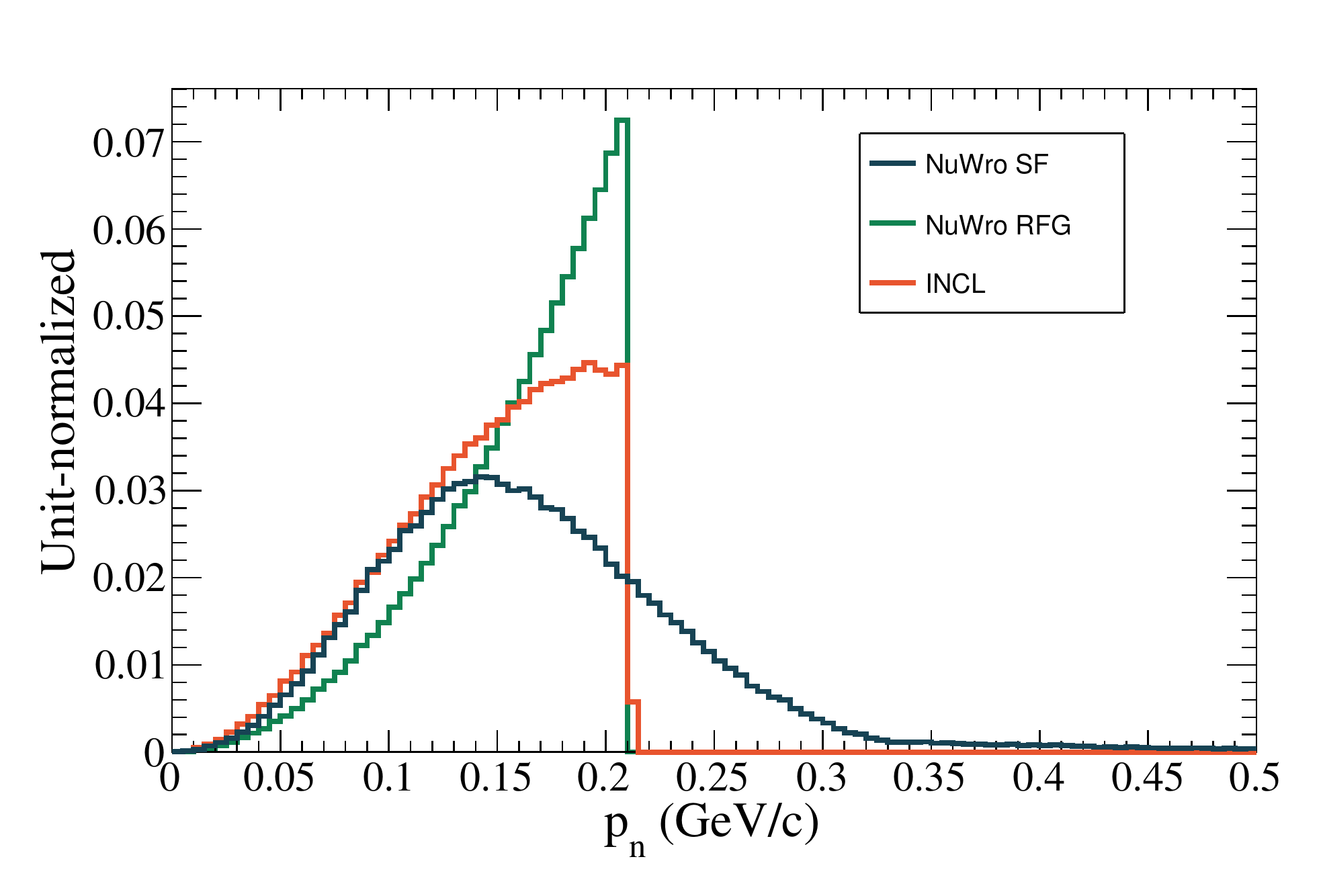}
\caption{\label{fig:n_mom} Momentum distribution of neutrons in the nucleus for NuWro SF and RFG and INCL nuclear models (shape comparison).}
\end{figure}

\begin{figure}[ht]
\centering
\begin{minipage}[h]{0.98\linewidth}
\center{\includegraphics[width=1\linewidth]{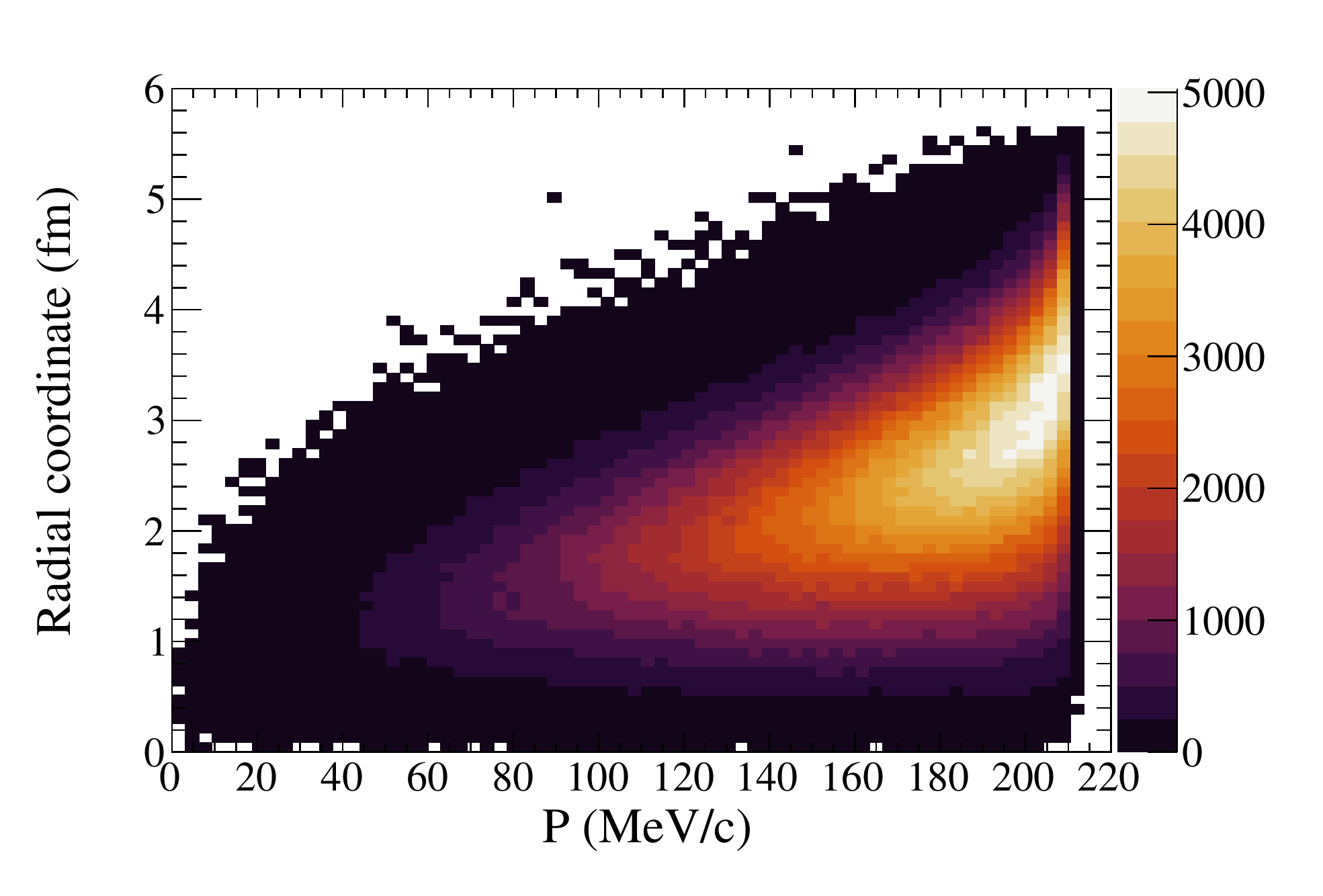} \\}
\end{minipage}
\hfill
\begin{minipage}[h]{0.98\linewidth}
\center{\hspace{-0.2cm}\includegraphics[width=0.95\linewidth]{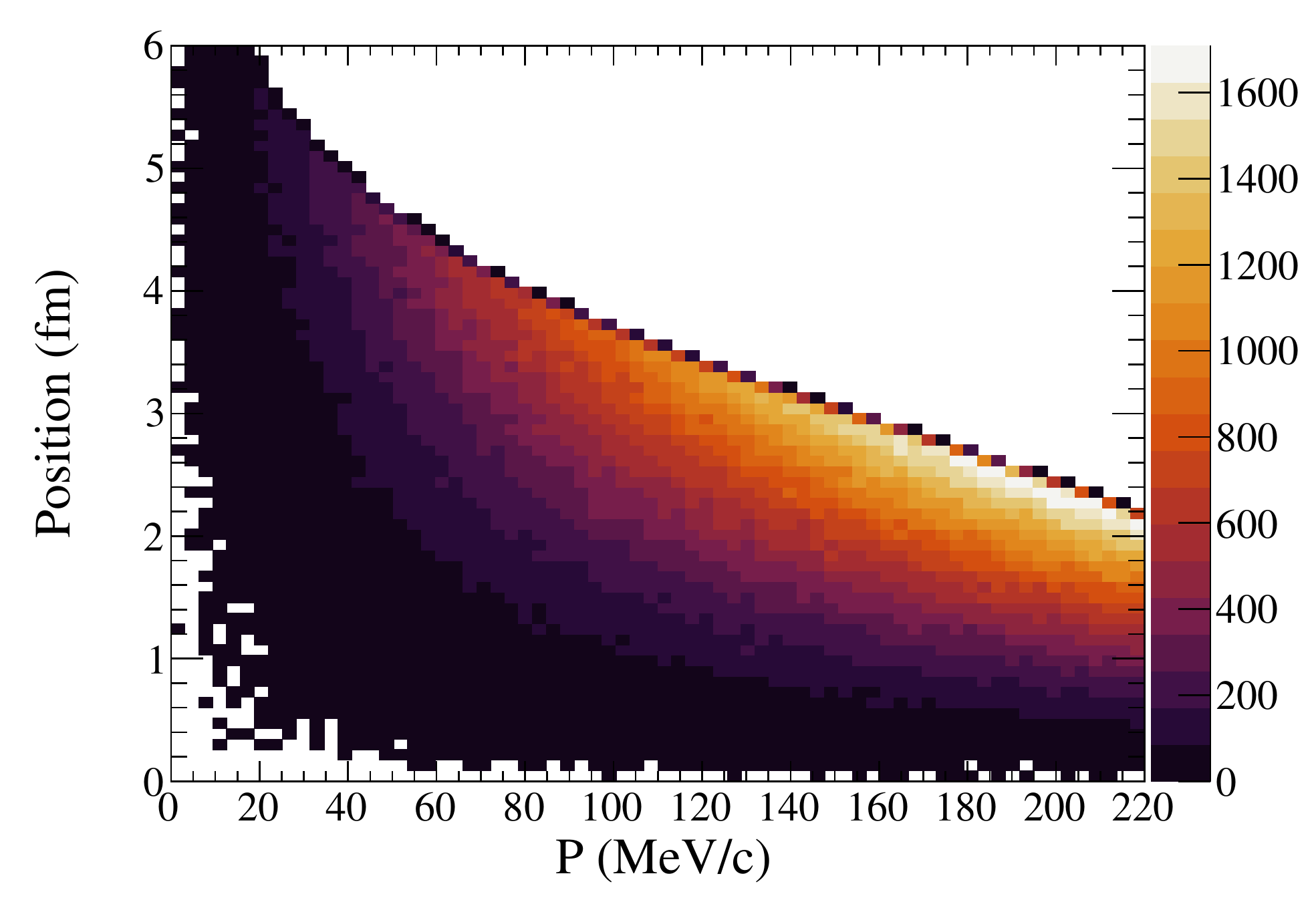} \\}
\end{minipage}
\caption{\label{fig:mom_pos1} Radial coordinate and momentum distribution inside the nucleus for INCL (top), and NuWro LFG (bottom) nuclear models} (z axis in arbitrary units).
\end{figure}

\begin{figure}[ht!]
\centering 
\includegraphics[width=0.98\linewidth]{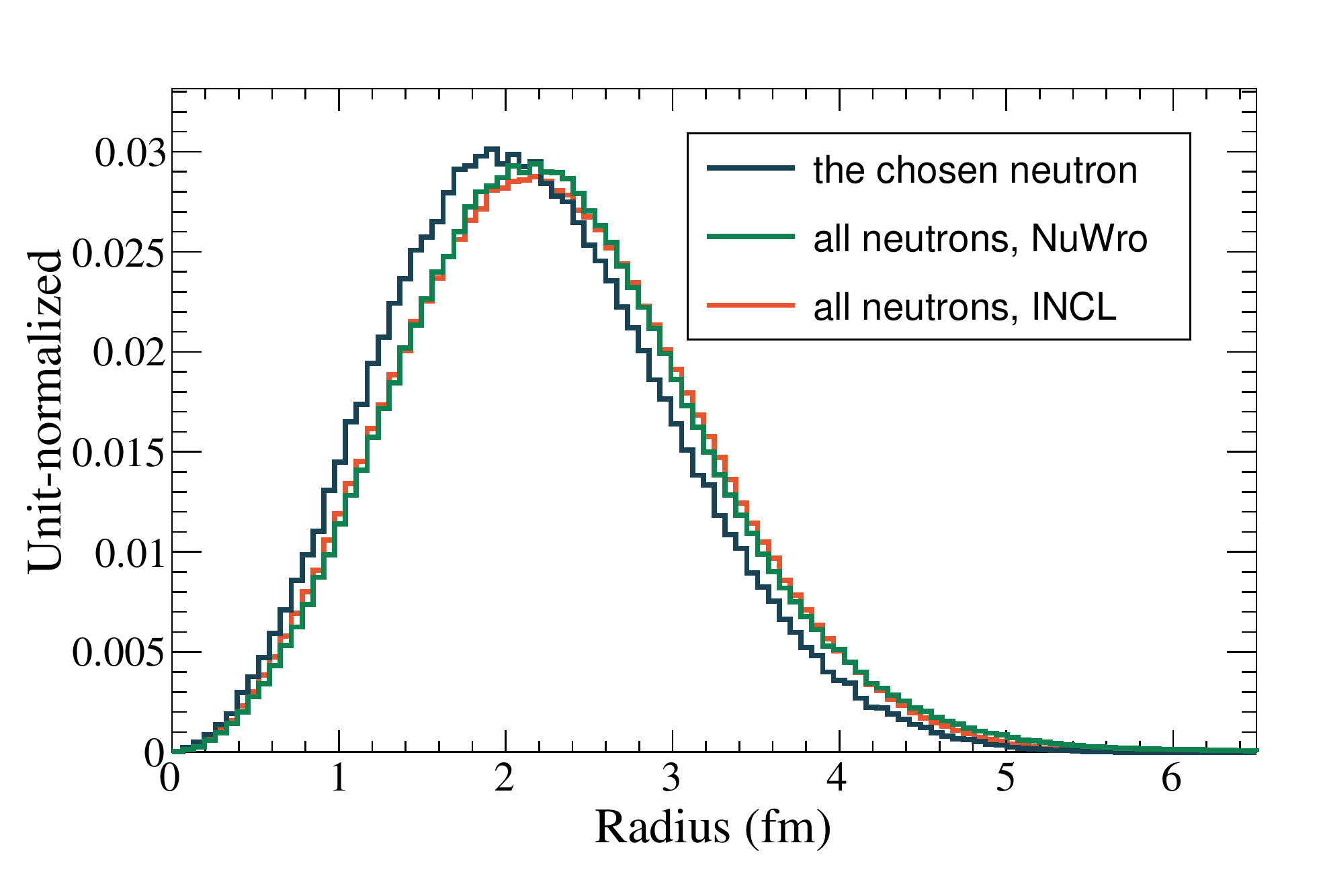}
\caption{\label{fig:n_mom2} Position distribution of neutrons in INCL and NuWro and of the INCL neutron chosen to match the SF neutron simulated in NuWro in our matching algorithm.} 

\end{figure}

Although driven by the same experimental charge density distribution, NuWro and INCL differ in their approaches to model the nucleus that can result in small inconsistencies within our algorithm. The momentum distribution of nucleons in SF and INCL is quite different, as can be seen in Fig.~\ref{fig:n_mom}: the high-momentum tail due to nucleon-nucleon correlations is missing in INCL (work is ongoing to include them, and we defer such results for a future paper).  Another discrepancy stems from the lack of correlation between position and momentum in the spectral function approach, while having the Hartree-Foch-Bogoliubov formalism implemented in INCL, as presented in Fig.~\ref{fig:mom_pos1}.
Thus, the INCL position distribution sampled starting from the simulated NuWro interactions is slightly different than the intrinsic nucleon distribution in INCL, as shown in Fig.~\ref{fig:n_mom2}.
In general, neutrons with high momentum in the SF tail tend to be attributed to the peripheral region of the nucleus, according to the INCL nuclear model shown in Fig.~\ref{fig:mom_pos1}, while in the NuWro SF model, there is no intrinsic position-momentum correlation for the target neutrons.
Still, as shown in Fig.~\ref{fig:n_mom2}, the chosen INCL neutron tends to be at smaller radius with respect to the general neutron distribution in INCL: this is due to the fact that the momentum-vector (and not the momentum magnitude) is used to find the INCL neutron which best matches the NuWro neutron.

It is worth mentioning that NuWro uses the local Fermi gas for nuclear model in the cascade itself. This model does include a momentum-position correlation, which is presented at the bottom of Fig.~\ref{fig:mom_pos1}. One can see that there exist a contrary dependence relative to the INCL solution, which is an important point in comparing these two cascades regardless of the primary vertex simulation.

In order to check that the results on FSI characterization are robust against the approximations of the simulation procedure above and, more in general, against assumptions on the initial state nuclear model, similar studies are also performed using Relativistic Fermi Gas and using a simulation representing the INCL nuclear model in the initial state. This is discussed in Appendix~\ref{app:pn}.

In the factorized models considered here, the FSI interaction cannot change the fundamental neutrino-nucleus interaction cross section. Thus for all studies presented here the NuWro cross section from SF (or from RFG for the results in Appendix~\ref{app:pn}) is considered.

The procedure to match NuWro and INCL models described above has been made possible by the high-level of modularity of Monte Carlo  implementations. It also opens the road to possible further improvement of FSI modeling into such simulation programs.

\section{Analysis strategy}
\label{sec:analysis}
In this section, the analysis strategy to compare and characterize the FSI effects in INCL and NuWro models is described.

Various channels are possible, depending on the particles leaving the nucleus: their probability is quantified in the two models. The FSI effects on the leading proton are then characterized by comparing its kinematics before and after FSI. Additionally, the Single Transverse Variables (STV), as introduced in Ref.~\cite{Lu:2015tcr}, are studied and compared between the two models. Resolution effects or thresholds are not applied in the results of Sec.~\ref{subsec:leadprot} and some of the studied variables are not observable experimentally. Comparison to data is deferred to Section~\ref{sec:data}, where direct comparison to STV measurements from T2K~\cite{T2K:2018rnz} and MINERvA~\cite{MINERvA:2018hba} are shown. In this case, acceptance cuts are applied to match the phase space covered by the two experiments.

\begin{figure}
    \centering
\includegraphics[width=0.98\linewidth]{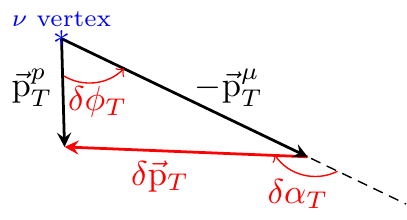}
    \caption{STV representation in the plane perpendicular to the neutrino direction.}
    \label{fig:stv_fig}
\end{figure}

The study focuses on the following STV:
\begin{equation}
\begin{gathered}
   \delta \alpha_{T} = arccos\frac{-\vec{p^{\mu}}_{T} \cdot \delta \vec{p}_{T}}{p^{\mu}_{T} \cdot \delta p_{T}} \\
   |\delta \vec{p_{T}}| = |\vec{p^{p}}_{T} + \vec{p^{\mu}}_{T}| \\
\end{gathered}
\end{equation}
where $\vec{p^{p}}_{T}$ is the component of the proton momentum projected into the plane transverse to the neutrino direction (transverse component) and $\vec{p^{\mu}}_{T}$ is the transverse component of the muon momentum.
The illustration of the STV definition is presented in Fig.~\ref{fig:stv_fig}. 
The variable $\delta \vec{p_{T}}$ could be considered as the ``missing transverse momentum'' and, in absence of FSI in quasi-elastic events, it represents the Fermi motion of the initial nucleon.
The variable $\delta \alpha_{T}$ is especially sensitive to the FSI of the leading proton. In transparent events (where the proton leaves the nucleus without FSI), the $\delta \alpha_{T}$ distribution depends only on the stochastic Fermi motion of the initial neutron and therefore is expected to be uniform. FSI tends to decrease the proton transverse momentum, inducing larger values of missing transverse momentum and $\delta \alpha_{T}$. 

The production of clusters is an important novelty brought by INCL: no other cascade model used within neutrino studies is able to simulate cluster production. Not only the kinematics of the simulated clusters but also their identification probability in scintillating detectors is studied.
To simulate clusters' interactions inside a detector, a simple Geant4 model~\cite{AGOSTINELLI2003250, Geant4, Allison:2016lfl}\footnote{For electromagnetic processes the standard physics list was used, for hadron processes we used  G4HadronPhysicsINCLXX. Additionally, G4DecayPhysics and G4IonINCLXXPhysics were used.} is built. A uniform, fully active block of hydrocarbon (with 1.06 g/cm$^3$ density corresponding to polystyrene) with dimensions big enough to contain a whole track (1900$\times$1900$\times$3000~cm$^3$ size) is simulated.  
The clusters are generated with momentum distribution as simulated by INCL FSI production. 

\begin{center}
\begin{figure}[ht!]
\includegraphics[width=0.98\linewidth]{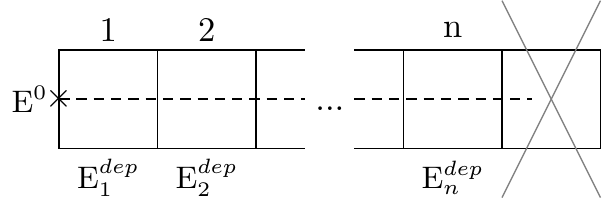}
\caption{Scheme of the particles simulation and identification algorithm in a segmented scintillating detector. The shaded line indicates the cluster track traversing multiple scintillating cubes. The cluster has initial kinetic energy $E^0$ and deposits energy $E^{dep}_i$ in each cube. The last cube, where the cluster stops, is not considered in the identification algorithm.}
\label{fig:pid_scheme}
\end{figure}
\end{center}

A simplified analysis is performed to gauge the observability of clusters: an identification algorithm is developed based on the path traversed by clusters and the energy deposited in the detector, as shown in the simple schematic of Fig.~\ref{fig:pid_scheme}.  

 The main idea of this algorithm is to try to identify the type of the particle based on its ionization curve. 
The measured deposited energy is used as an estimation of the particle momentum along its trajectory. The measured local deposits of energy ($dE/dx$) along the track are compared to the expected ones for different clusters \{$\alpha$, D, T, $^3$He, p\} at the estimated momenta. Finally the algorithm selects the cluster hypothesis that best matches the observed $dE/dx$ along the track.

Actual identification capabilities in an experiment depend highly on the detector geometry, granularity, and scintillator material. Here we consider a detector granularity of 1~cm$^3$ cubes, corresponding, for instance, to the geometry of the ND280 upgrade scintillating target superFGD~\cite{T2K:2019bbb}. We do not take into account any reflecting material or border effects.

Tracks are simulated as straight lines parallel to one of the cube faces and starting from the cube boundary. The last part of the track, which is shorter than 1~cm, is not used in the analysis since the exact length of the last step is unknown, given the simulated granularity. Thus we effectively remove the Bragg peak from this simplified identification algorithm. In more sophisticated analyses, the Bragg peak study could bring further information to help particle identification. Moreover, at the end of the track, a fraction of events undergoes inelastic interactions with the creation of secondary particles. We leave the investigation of such secondary interactions and their detector signature for future work: we do not include the energy of secondary particles in the analysis since we remove the last cube. 

We consider only the energy deposited by the primary particle by ionization and we apply Birks quenching. Depending on the material, an additional overall suppression factor should be considered in energy to scintillation light conversion but is omitted in the present study.

Visible energy in a given simulation step is therefore calculated as
\begin{equation}
    \mathrm{E}^{vis}_{step} = \frac{\mathrm{E}^{dep}_{step}}{1 + k_B \cdot \frac{\mathrm{E}^{dep}_{step}}{\mathrm{L}_{step}}}.
\end{equation}
where $\mathrm{E}^{dep}_{step}$ is the energy deposited in the step, k$_B$ is the Birks coefficient, and $\mathrm{L}_{step}$ is the step length.
Birks coefficient for protons is taken from Ref.~\cite{T2KND280FGD:2012umz}, where $k_B$ = 0.0208~(cm/MeV); according to Ref.~\cite{GSO}, the coefficients for protons and deuterons are assumed to be the same. The same Birks coefficient is also assumed for tritium. For $\alpha$ particles the results from Ref.~\cite{Sarazin:2012ppa} are used, where $k_B$ = 0.0085~(cm/MeV).  

The total visible energy is evaluated as the sum of the visible energy in each cube is
\begin{equation}
E^{vis} = \sum_{i=1}^{n} E^{vis}_i.
\end{equation}
This visible energy is a proxy for the total kinetic energy of the particle. However, it is not a perfect estimator since the energy lost in the last cube, where inelastic events may happen, is discarded from the analysis. As an analogous estimator of the remaining kinetic energy of the cluster at each step along the track, the total `residual' visible energy is estimated as
\begin{equation}
\label{eq:evis}
E^{vis, res}_i = E^{vis} - \sum_{m=1}^{i} E^{vis}_m.
\end{equation}
The distribution of the visible energy deposited in each step $E^{vis}_i$, as a function of the left visible energy $E^{vis, res}_i$, is used to build the expectations for each of the five particle hypotheses  \{$\alpha$, D, T, $^3$He, p\}. For each particle $k$, the mean of the simulated visible energy at a given step is estimated ($E^{exp,k}_i$) for the corresponding value of remaining visible energy ($E^{vis, res}_i$). The width of such distribution is also evaluated ($\sigma^k_i$).
Finally for each step of the simulated cluster track, the visible energy is compared to the expected one for each particle hypothesis: a $\chi^2$ is built for each particle hypothesis
\begin{equation}
\chi_k^2= \sum_{i=1}^{n} \frac{(E^{vis}_i - E^{exp,k}_i)^2}{(\sigma^{k}_i)^2}
\end{equation}
and the hypothesis with lowest $\chi^2$ is chosen to identify the particle.

\section{Results}
\label{sec:res}

In Tab.~\ref{tab:channelsSF} the fraction of different final states produced by INCL and NuWro FSI simulation from CCQE interactions are reported. 
 In the NuWro SF simulation, due to short-range correlations, the neutrino vertex contains two outgoing protons in 15\% of events.
In the INCL vertex, only the leading proton, defined as the proton with higher kinetic energy, is retained and triggers the cascade.
 We have also tested that complete removal of events with 2-protons  at the neutrino vertex does not impact the conclusions on the characterization of the FSI cascade of the leading proton in NuWro and INCL.  We consider only the leading proton and its secondary hadrons for the channel characterization in Tab.~\ref{tab:channelsSF}. If the proton has the same energy before and after FSI, the event is characterized as ``no FSI'' (a transparent event). 
 
 \begin{table}[htbp]
\centering
\caption{\label{tab:channelsSF}  Fractions of the different FSI channels, i.e. fraction of events with different final state particles after the FSI cascade, in CCQE events with T2K neutrino energy flux. Fractions of events with and without protons in the final state are quoted separately, while in the text the percentages with respect to the total number of events are quoted. }
\begin{tabular}{|c|c|c|c|}
\hline
&\multirow{2}{*}{\textbf{Channel}} & \multirow{2}{*}{\textbf{NuWro SF}} & \textbf{INCL}\\
&&&\textbf{+NuWro SF}\\\hline
& no protons & 1.37\% &19.47\% \\
& protons & 98.63\% &80.53\%  \\
\hline
\hline
\multicolumn{1}{|c|}{\multirow{4}{*}{\rotatebox{90}{\textbf{no proton}}}}& absorption  & 4.45\% & 39.49\% \\ 
\multicolumn{1}{|c|}{}&neutron + $\pi$ production  & 3.40\%  &0.60\% \\  
\multicolumn{1}{|c|}{}&$\pi$ production& 0.21\% & 0\% \\

\multicolumn{1}{|c|}{}&neutron knock-out& 91.4\% &29.58\% \\
\multicolumn{1}{|c|}{}&cluster knock-out & 0\%  & 30.33\% \\
\hline
\hline
\multicolumn{1}{|c|}{\multirow{4}{*}{\rotatebox{90}{\textbf{proton}}}}& 1 proton, no FSI   & 70.38\% & 68.49\% \\ 
\multicolumn{1}{|c|}{}&1 proton only with FSI  & 2.45\%  &19.21\% \\  
\multicolumn{1}{|c|}{}&1p + other nucleons or clusters& 26.21\% &11.68\% \\
\multicolumn{1}{|c|}{}&proton(s)+ $\pi$ production  & 0.96\%  &0.62\% \\
\hline
\hline
\end{tabular}
\end{table}

INCL features an evident enhancement of events with absorption of the proton and only a muon in the final state: 
7.7\% of the total events in INCL against less than 0.1\% in NuWro. 
Indeed, in the NuWro cascade, the vast majority of particles produced, and not Pauli blocked, leave the nucleus. The INCL nuclear model features larger probability of reabsorbing particles in the nucleus during FSI. This tendency is also confirmed by the smaller fraction of events with more than 1 proton in the final state (multinucleon production by FSI results into 25.6\%  of total events in NuWro and 9\% in INCL). 

\begin{figure}[ht]
\centering
\includegraphics[width=0.98\linewidth]{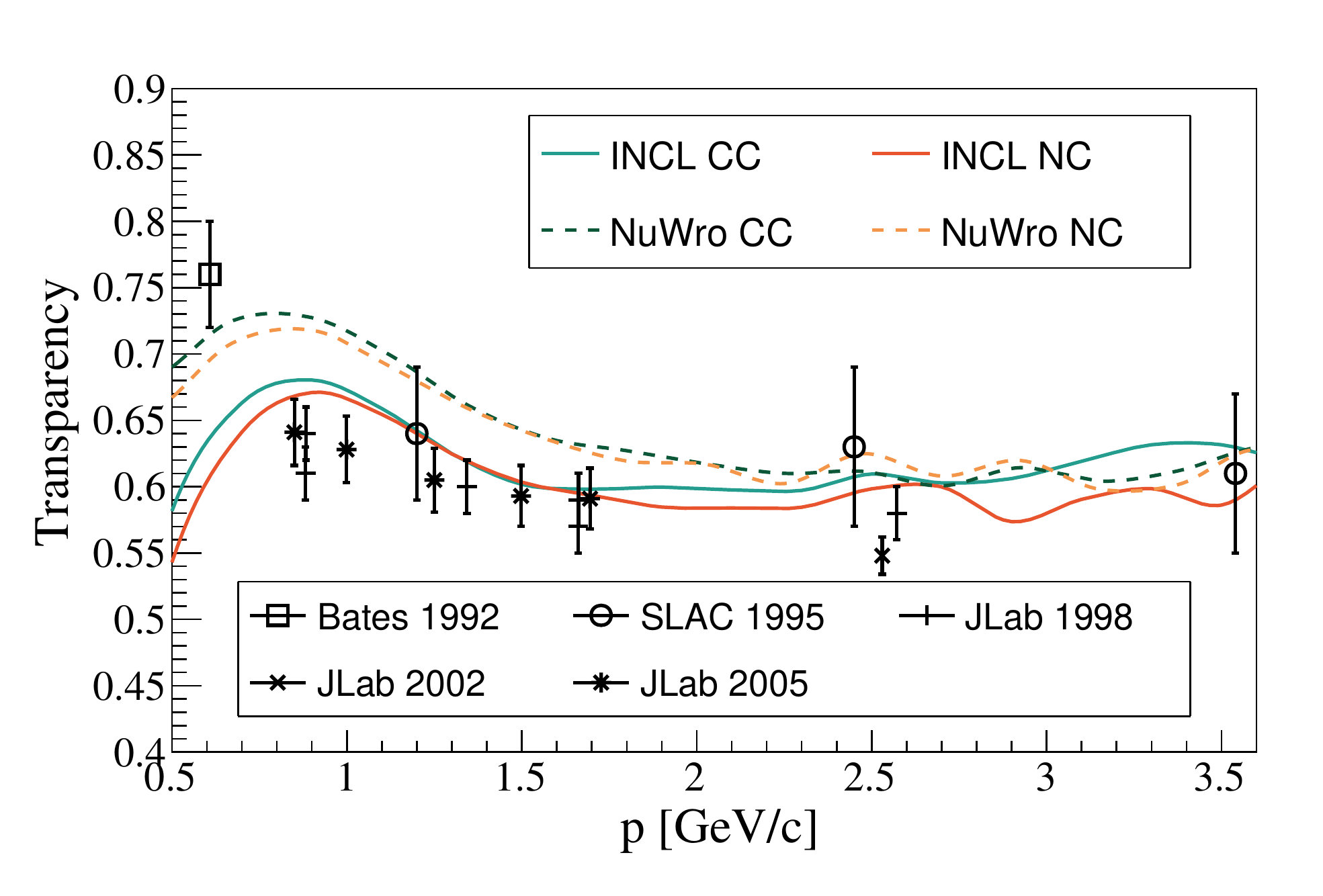}
\caption{\label{fig:transp} Nuclear transparency of $^{12}$C (percentage of events without FSI) as a function of proton momentum modelled by NuWro and INCL.}
\end{figure}

 We have also compared both INCL and NuWro cascades against existing transparency data. Here we define transparency as the percentage of events without FSI interaction on the leading proton exiting from the interaction vertex. Fig.~\ref{fig:transp} shows that both models are in good agreement with data while having a divergence in their prediction. Especially for the low momentum protons, INCL features smaller transparency. The difference between the two models is dominated by the impact of the short-range correlations on the FSI definition, as also shown in Refs.~\cite{Niewczas:2019fro} and~\cite{PhysRevD.104.053006}. In the region of interest for the present study ($0.2-1$~GeV of proton momentum), the predictions vary the most, while there are very sparse data and no clear preference for one model over the other. 

The larger FSI strength in INCL suggests a larger dissipation of energy across the nucleus through interactions that tend to be more `soft'. 
On the other hand, the nuclear model of INCL includes the probability to form  clusters during the attempt of the nucleon to leave the nucleus, as explained in Sec.~\ref{sec:sim}.
 Indeed, the events with no proton in the final state are in the large majority due to charge exchange in NuWro (91\% of neutron production) while in INCL the probability of cluster production and neutron production in events without protons in the final state is similar (around 30\% each).

\subsection{Leading proton kinematics}
\label{subsec:leadprot}
\begin{figure*}[h!]
\centering
\begin{minipage}[h]{0.49\linewidth}
\center{\includegraphics[width=1\linewidth]{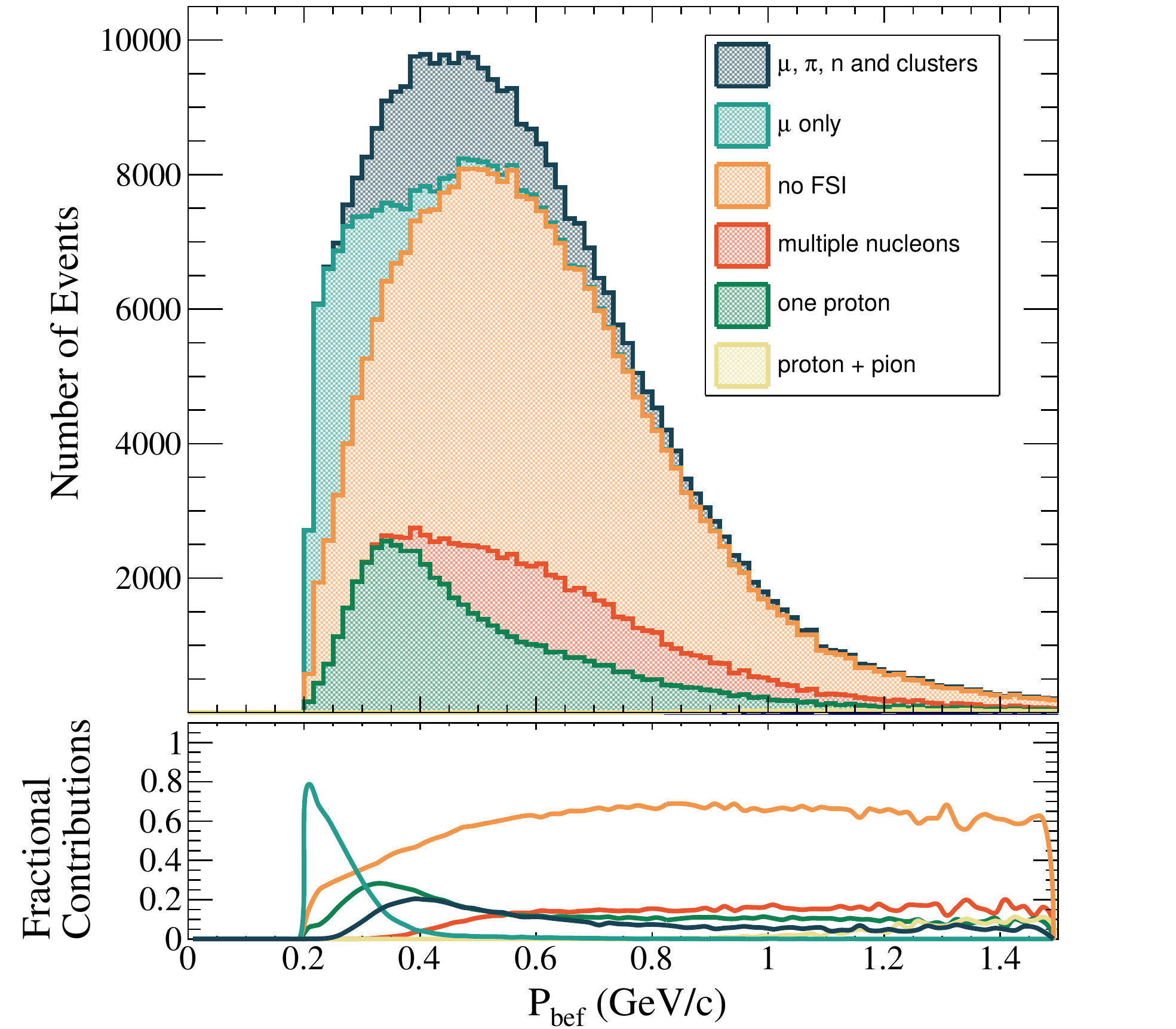} \\}
\end{minipage}
\hfill
\begin{minipage}[h]{0.49\linewidth}
\center{\includegraphics[width=1\linewidth]{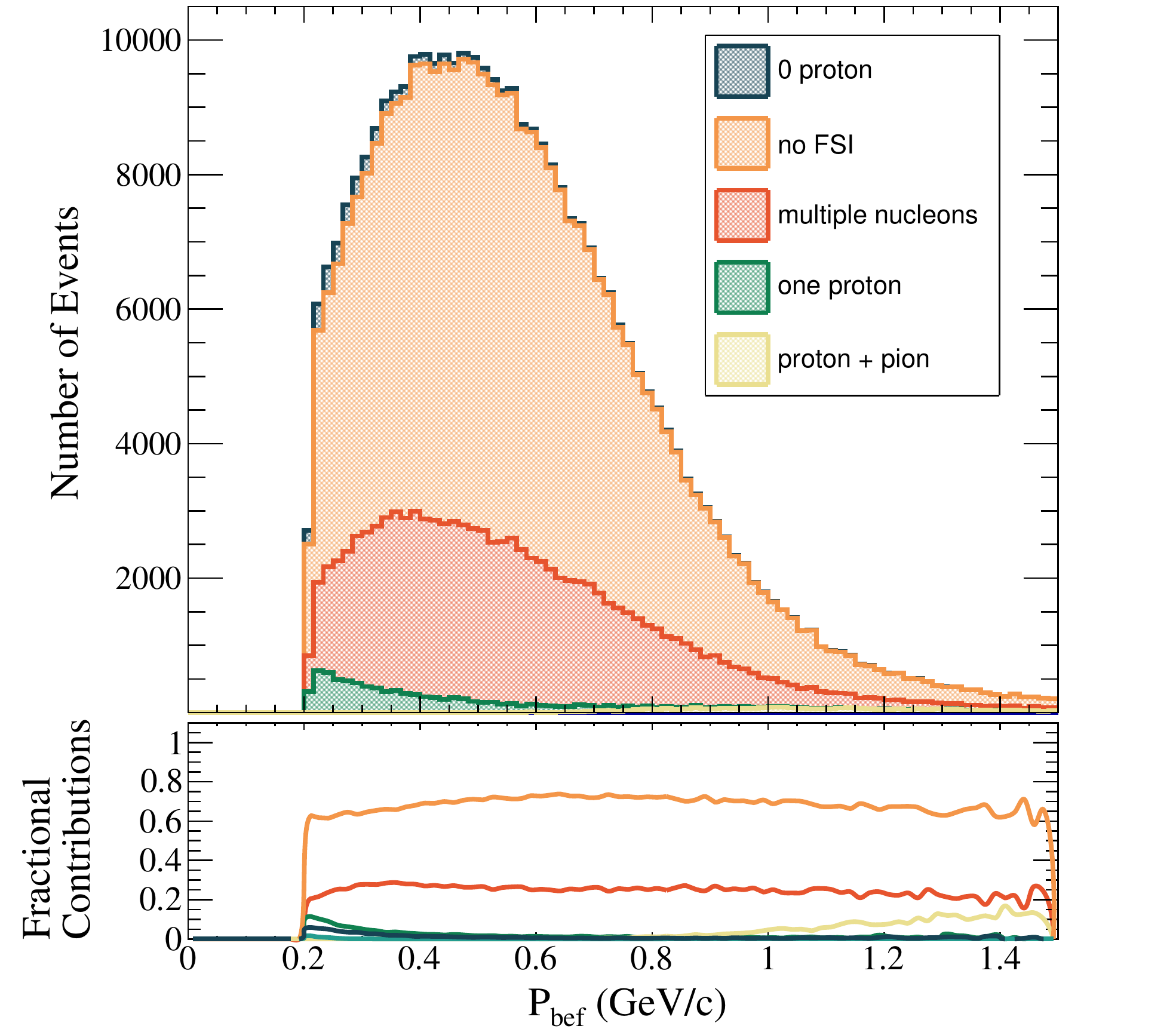} \\}
\end{minipage}
\vfill
\begin{minipage}[h]{0.49\linewidth}
\center{\includegraphics[width=1\linewidth]{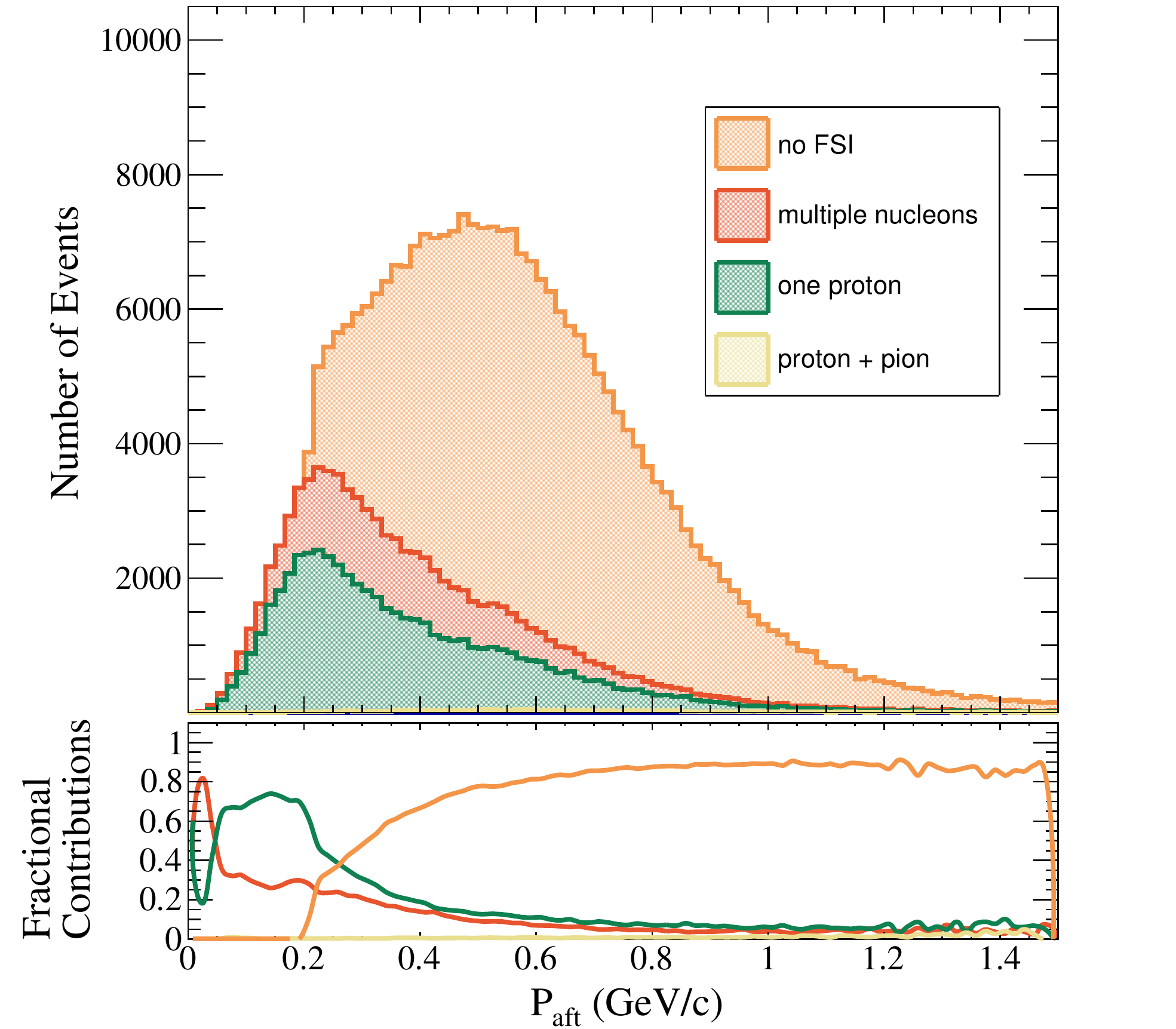} \\}
\end{minipage}
\hfill
\begin{minipage}[h]{0.49\linewidth}
\center{\includegraphics[width=1\linewidth]{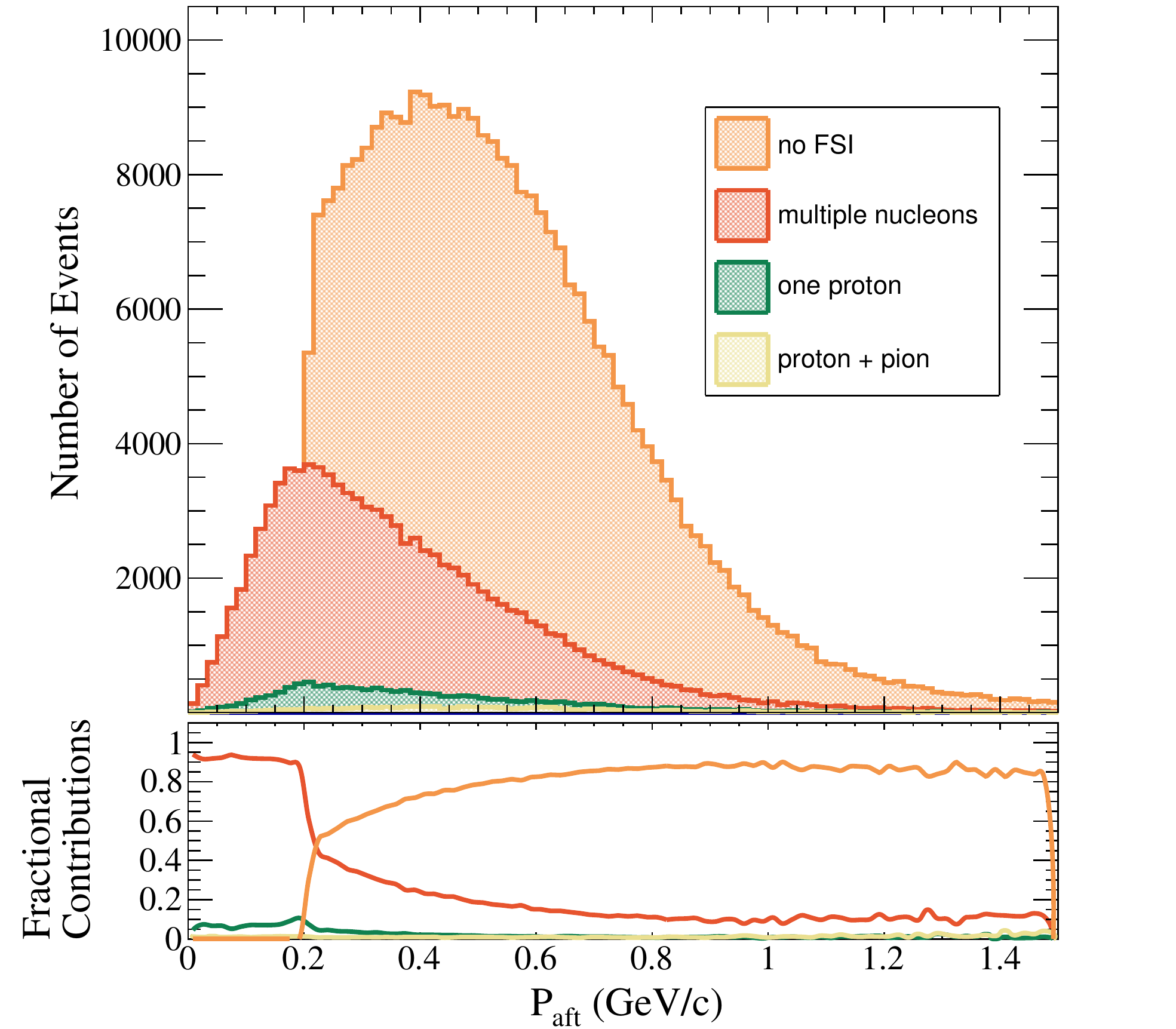} \\}
\end{minipage}
\caption{\label{fig:mom_bef} Top: proton momentum before FSI for INCL (left), and NuWro (right) cascade models in CCQE events with T2K neutrino energy flux.  Sub-processes correspond to the fate of the proton after FSI. The shape of proton momentum before FSI is by definition  identical for INCL and NuWro cascades. Bottom: leading proton momentum after FSI for INCL (left), and NuWro SF (right) nuclear models. The fractions of different FSI sub-processes as listed in Tab.~\ref{tab:channelsSF} is also shown. The 0 proton channel in NuWro includes muon only and pion and neutron production. There is no cluster production in NuWro.}
\end{figure*}

The leading proton momentum before and after FSI is shown in Fig.~\ref{fig:mom_bef}. Events with the production of other particles, e.g., clusters, but no protons in the final state, are included only in the distribution of proton momentum before FSI. 
 The proton momentum before FSI has by construction the same distribution for NuWro and INCL, as explained in Sec.~\ref{sec:sim}. In Fig.~\ref{fig:mom_bef} the relative fraction of the different final states as a function of proton momentum before and after FSI is also shown. A larger diversity of channels is visible in the INCL case. As previously discussed, the smaller transparency of INCL induces a smaller number of events with at least one proton in the final state after FSI. In particular, all the events with full proton absorption are concentrated at low  proton momentum before FSI, while the events with cluster production populate the distribution just below the main peak of initial proton momentum (around 350~MeV/c). 
Again, due to the smaller nuclear transparency of INCL, the events where the proton leaves the nucleus are much less often accompanied by additional nucleons. Thus in NuWro, the protons at very low momenta after FSI are almost always accompanied by other nucleons, while in INCL, they are mostly events with only one proton.

\begin{figure*}[t!]
\centering
\begin{minipage}[h]{0.49\linewidth}
\center{\includegraphics[width=1\linewidth]{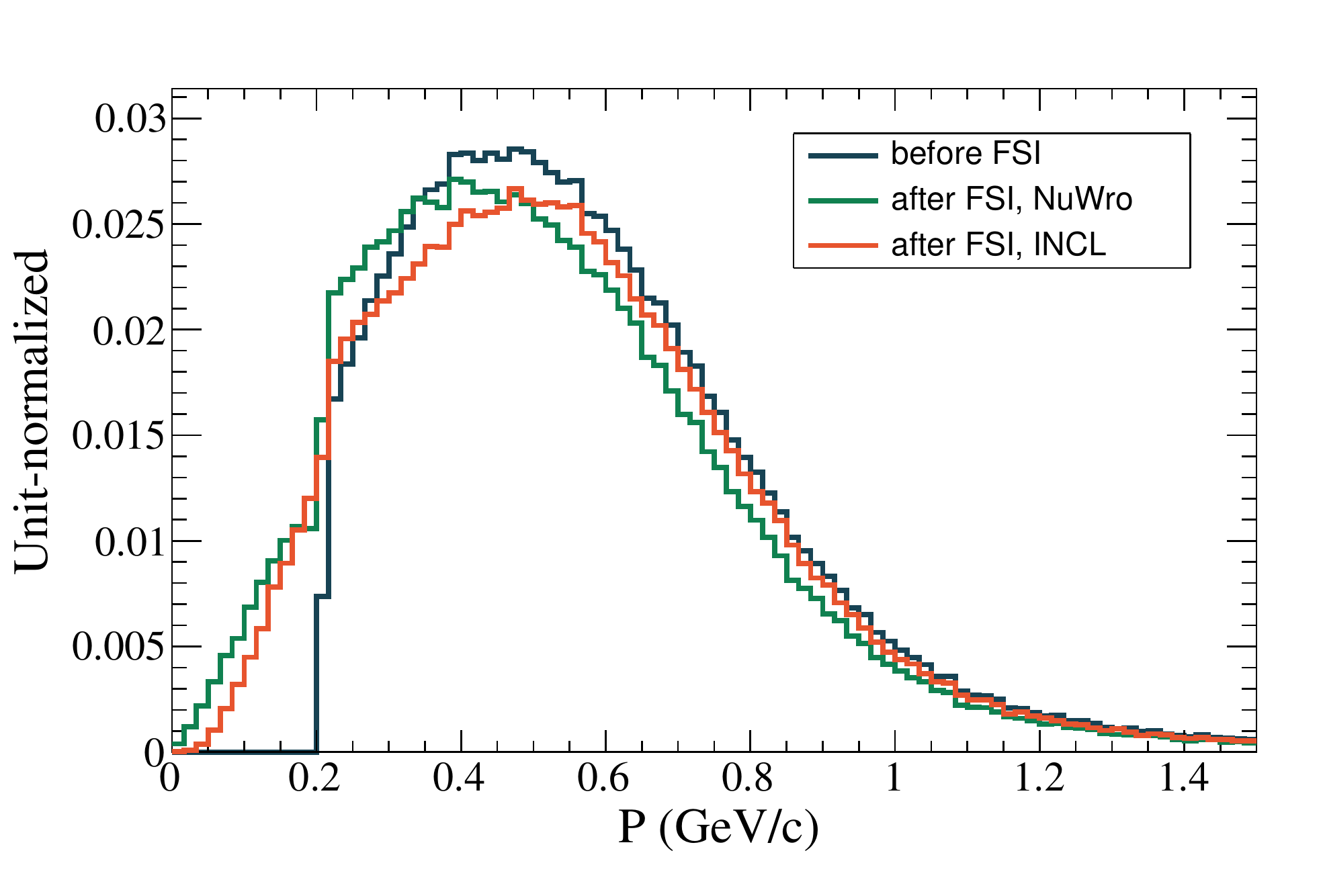} \\}
\end{minipage}
\hfill
\begin{minipage}[h]{0.49\linewidth}
\center{\includegraphics[width=1\linewidth]{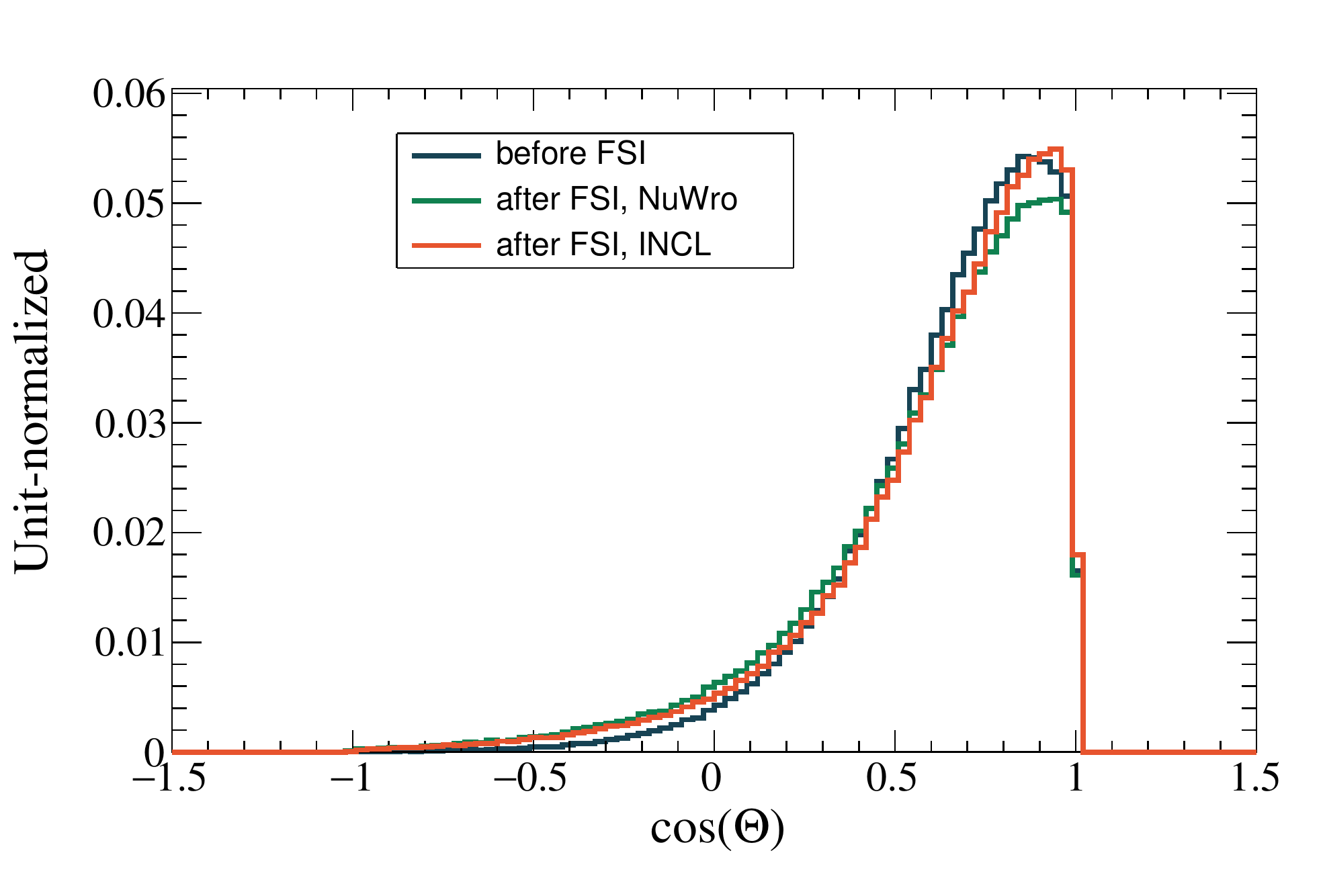} \\}
\end{minipage}
\vfill
\begin{minipage}[h]{0.49\linewidth}
\center{\includegraphics[width=1\linewidth]{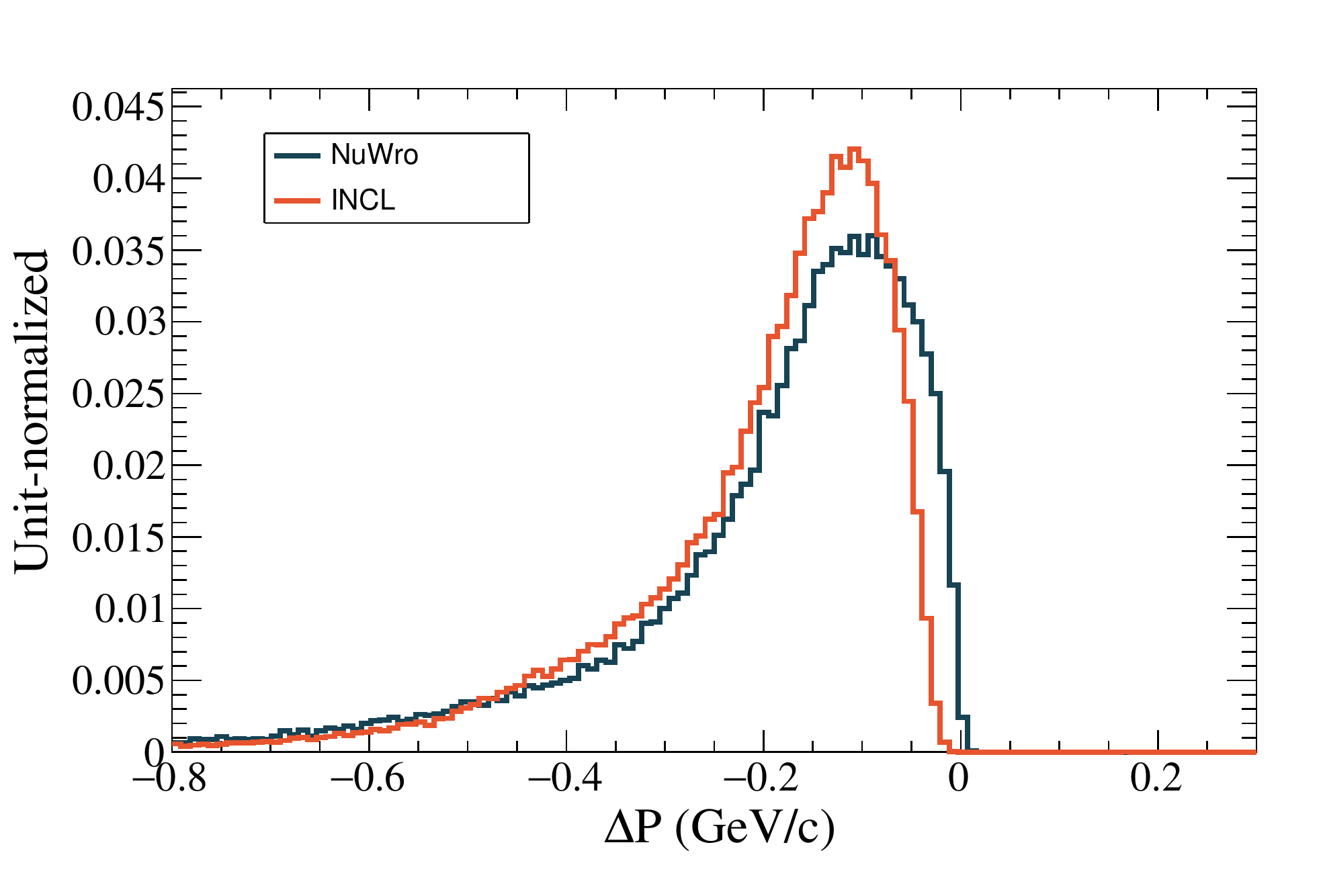} \\}
\end{minipage}
\hfill
\begin{minipage}[h]{0.49\linewidth}
\center{\includegraphics[width=1\linewidth]{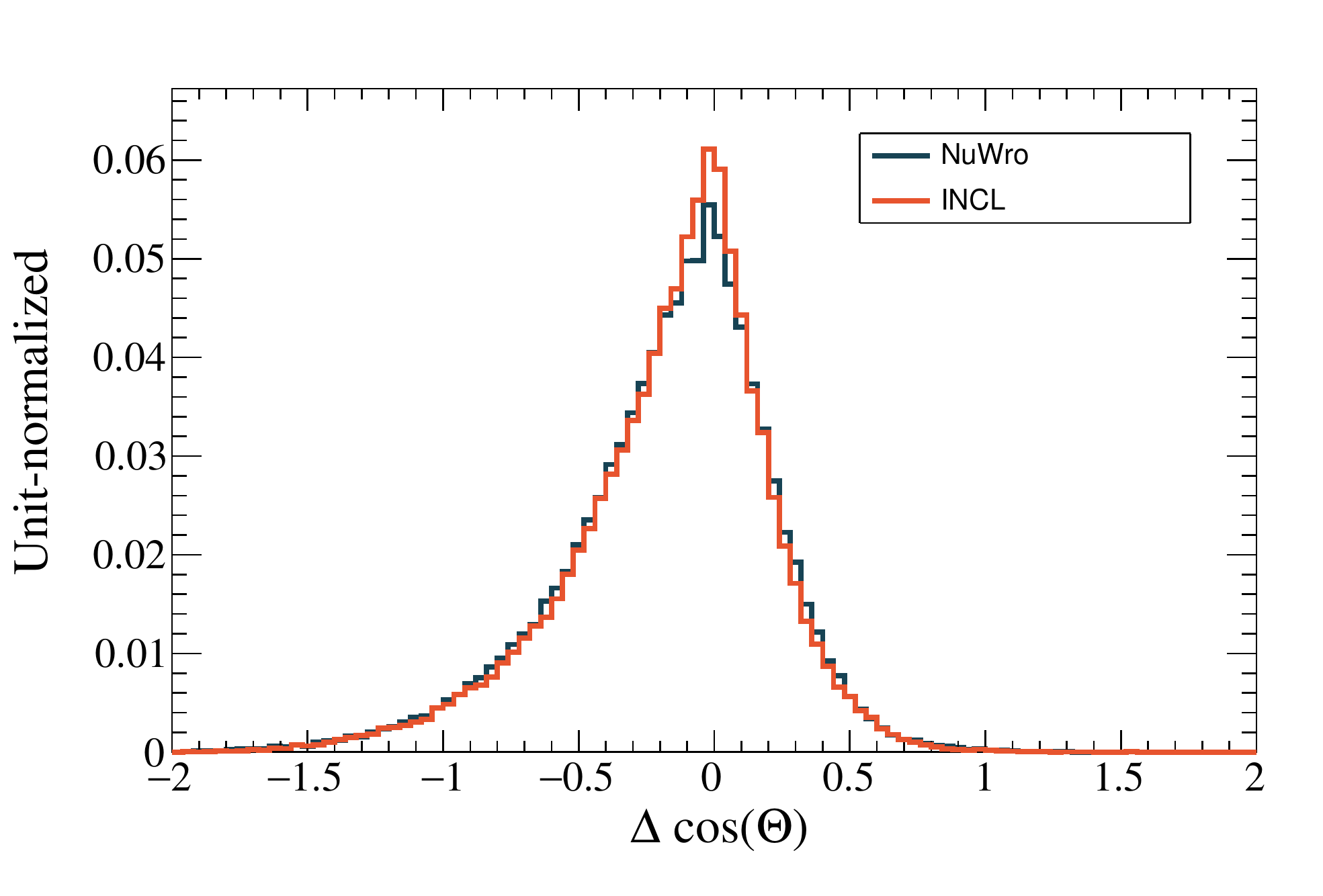} \\}
\end{minipage}
\caption{\label{fig:pdcosT_befaft} Distribution of leading proton's momentum in NuWro SF and NuWro+INCL before and after FSI (top left, shape comparison). Distribution of difference in proton's momentum before and after FSI ($\Delta P=P_{after} - P_{before}$) in NuWro SF and NuWro+INCL (bottom left, shape comparison). Distribution of leading proton $\cos(\Theta$) in NuWro SF and NuWro+INCL before and after FSI (top right, shape comparison). Distribution of difference in proton's $\cos(\Theta$) before and after FSI ($\Delta \cos\Theta =\cos\Theta_{after} - \cos\Theta_{before}$) in NuWro SF and NuWro+INCL (bottom left, shape comparison). The ``no FSI'' channel is excluded. CCQE events with T2K neutrino energy flux are simulated.}
\end{figure*}

More direct FSI quantification on the leading proton is done by comparing the kinematics of the proton before and after FSI, as shown in Fig.~\ref{fig:pdcosT_befaft}.
As expected, FSI decelerate the leading proton and induce smearing compared to the pre-FSI angular distribution that is peaked in the forward direction (but this effect is less evident in INCL). 
Interestingly, INCL tends, event-by-event, to decelerate  protons more but still the distribution of momenta of all the protons leaving the nucleus after FSI tends to be larger in INCL, since the low momenta protons are absorbed in the nucleus.

\begin{figure}[ht]
\centering
\includegraphics[width=0.98\linewidth]{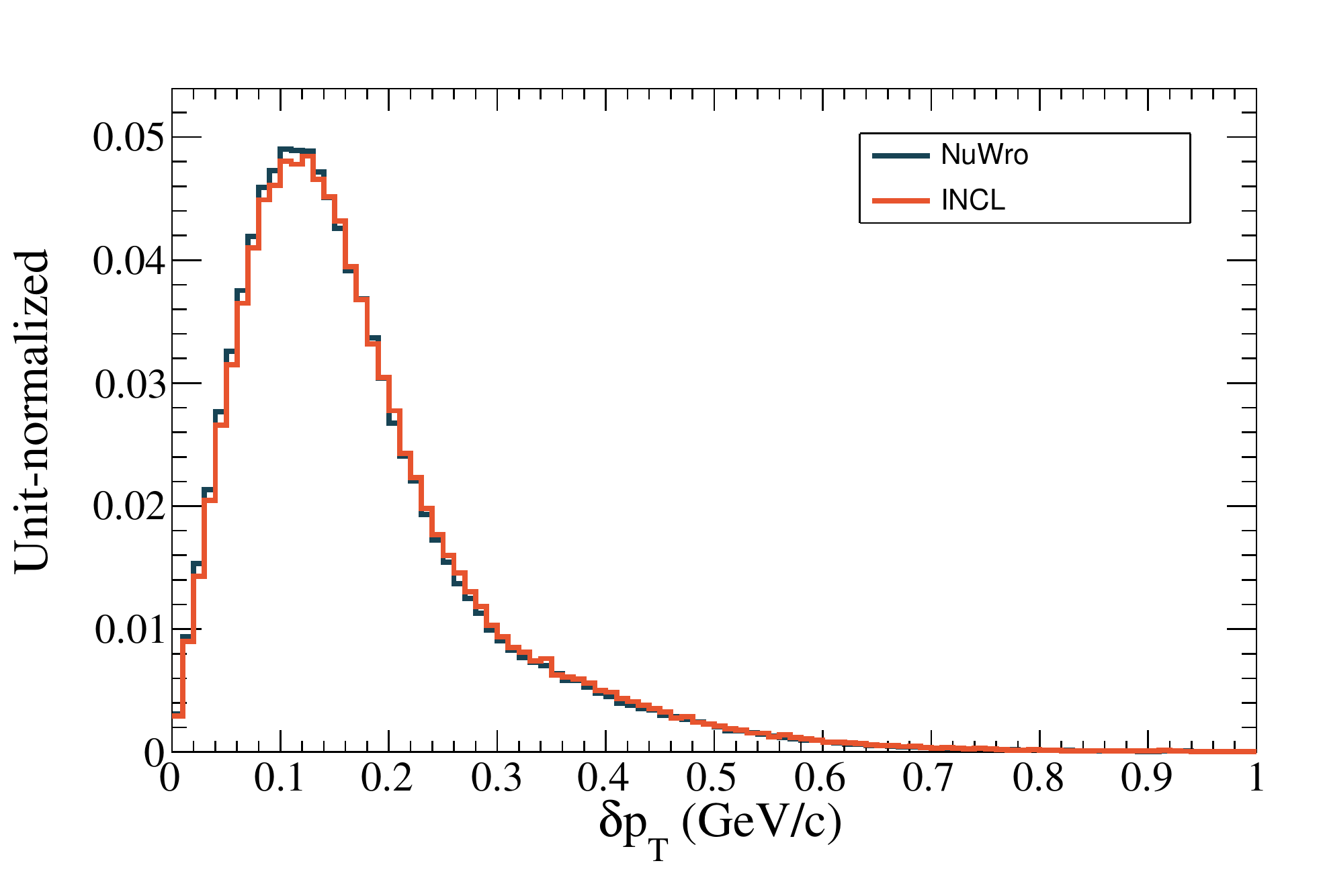}
\vfill
\includegraphics[width=0.98\linewidth]{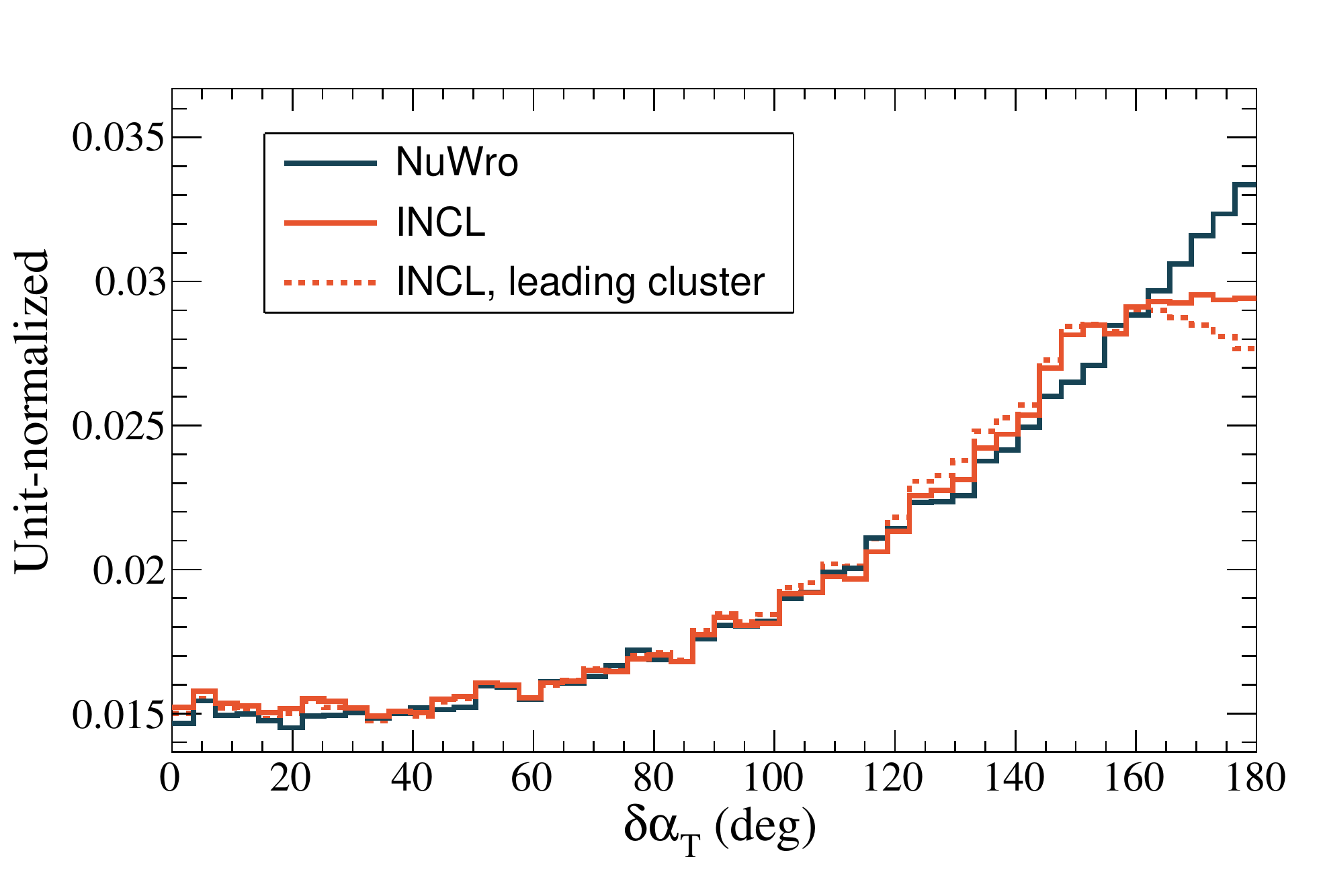}
\caption{\label{fig:dpt} Top: $\delta$p$_T$ distribution in NuWro SF and NuWro+INCL (shape comparison). Bottom:  $\delta\alpha_T$ distribution in NuWro SF and NuWro+INCL (shape comparison). The effect on  $\delta\alpha_T$ shape of including also the leading nuclear cluster in INCL events is shown. CCQE events with T2K neutrino energy flux are simulated.}
\end{figure}

In order to disentangle FSI effects from the physics of the initial nuclear state, STV are used. As explained in Sec.~\ref{sec:analysis}, the identified experimental observable most sensitive to FSI is $\delta\alpha_T$, while $\delta p_T$ is primarily sensitive to the initial nuclear state. This is confirmed by the results in Fig.~\ref{fig:dpt}, where such distributions are  compared between NuWro and INCL. The shape of $\delta p_T$ is very similar between the two models since the initial nuclear momentum is taken from the same model.
The most visible feature is the suppression of the large $\delta\alpha_T$ values in INCL, corresponding to a suppression of low momentum protons. This conclusion is robust, independently of the nuclear model used for the fundamental interaction: we observe a similar behavior using RFG, as shown in Appendix~\ref{app:pn} in Fig.~\ref{fig:rfg}.
Part of the low momentum protons lost in INCL are emitted as clusters. Such clusters will not have the same detector signature of protons, as discussed in the following, and they do not fill the suppression at very large values of $\delta\alpha_T$, as can be seen in Fig.~\ref{fig:dpt}: indeed, there is still a sizable fraction of events, at low proton momenta, with full proton absorption and only a muon in the final state.

\subsection{Nuclear cluster production}
\label{subsec:nucl_clus}
\begin{figure}[ht]
\centering
\includegraphics[width=0.98\linewidth]{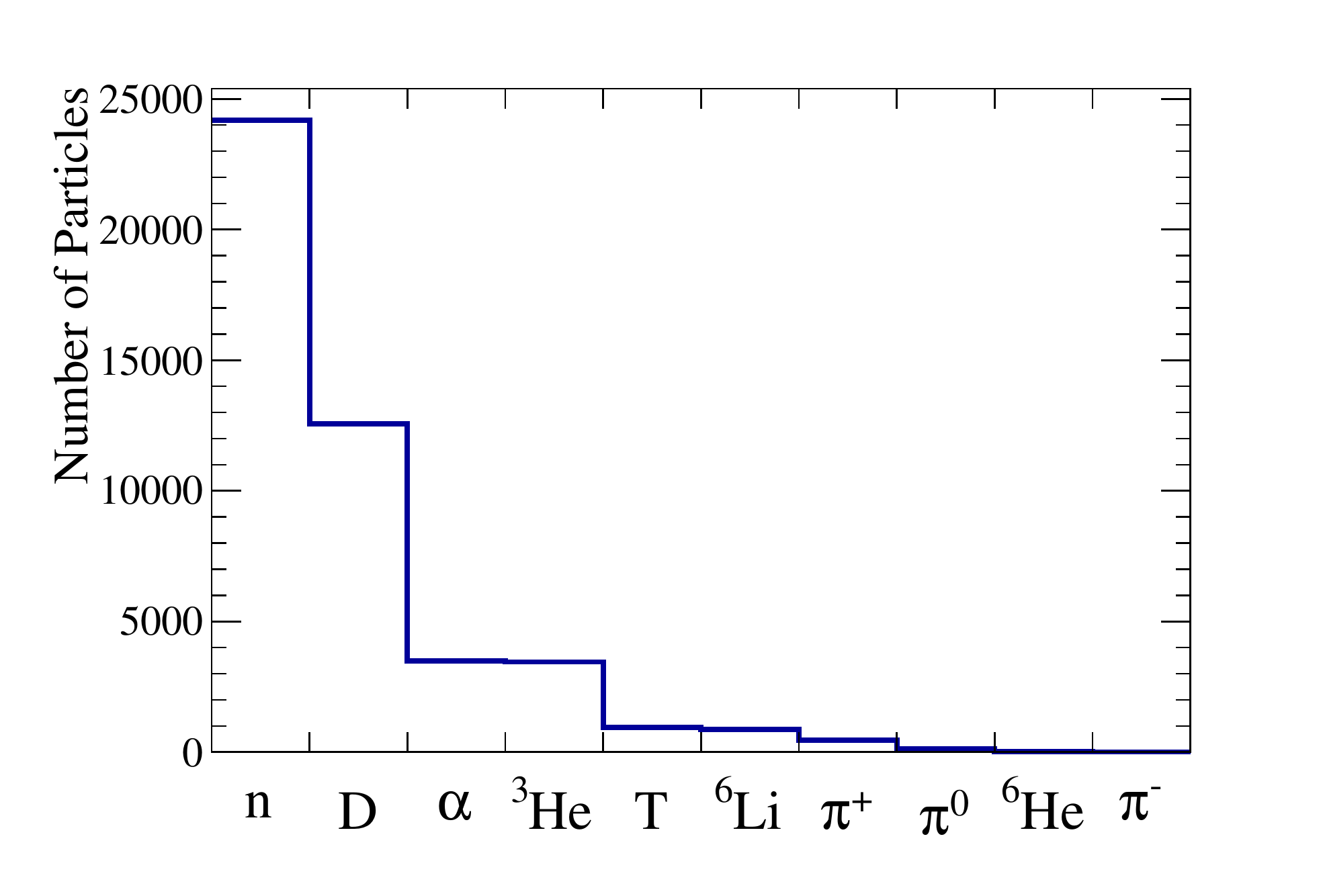}
\caption{\label{fig:part0p} Particles leaving the nucleus in events without proton in the final state in INCL.}
\end{figure}

In events with proton absorption {\and multiple nucleons production} in INCL, different particles can leave the nucleus, as shown in Fig.~\ref{fig:part0p}.  As one can see, INCL features a notable production of deuterons, $\alpha$-particles, $^3$He isotopes and tritons. In this paper, we focus on studying production of these nuclear clusters.
The production of clusters in nucleon scattering on nuclei is a well established evidence in nuclear physics, see for instance Ref.~\cite{neutron_clus}.

The production of clusters by FSI in neutrino interactions is a novelty of this study. While the presence of clusters can change the interpretation of various measurements in data (proton multiplicity, vertex activity, single transverse variables, etc.) the signature of such clusters is very detector-dependent, notably regarding granularity and threshold of measurable energy deposits. We provide here the fundamental information to characterize the observability and signature of such clusters in different detectors. In particular, the path length of cluster tracks, their mis-identification probabilities, evaluated with a simplified identification algorithm, and the energy released around the neutrino vertex are quantified.

\begin{figure}[ht]
\centering
\includegraphics[width=0.98\linewidth]{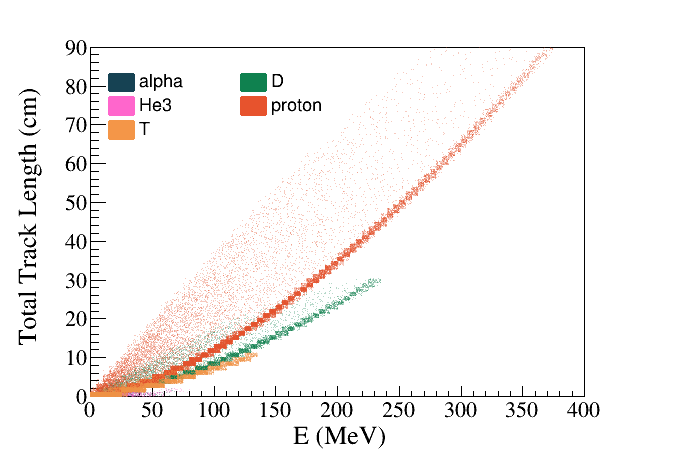}
\caption{\label{fig:e_step} Total track length of nuclear clusters as a function of visible energy of the particles. The $^3$He distribution overlaps with the $\alpha$ distribution.}
\end{figure}

\begin{figure}[h!]
\centering
\includegraphics[width=0.98\linewidth]{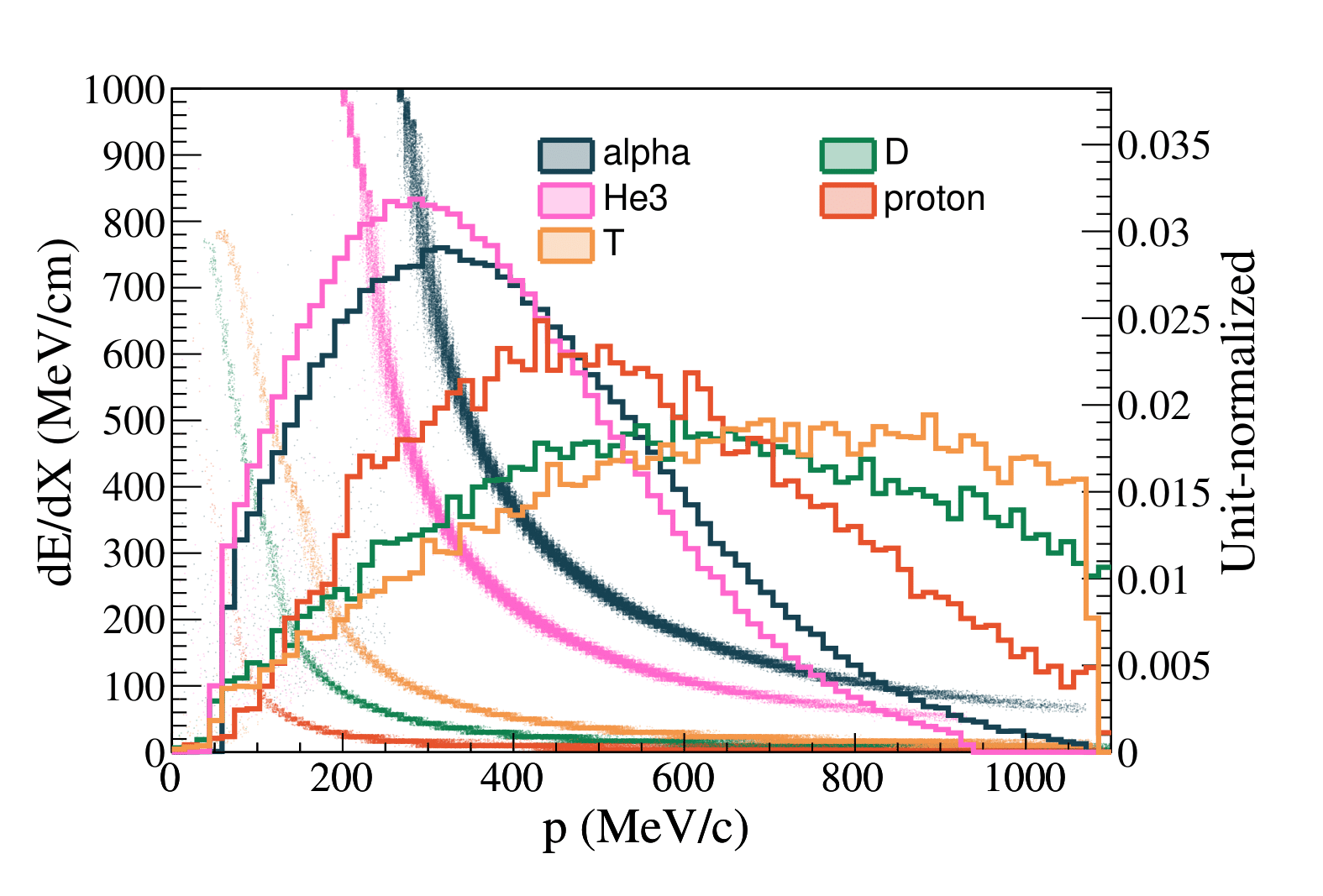}
\caption{\label{fig:dEdX_birks} Momentum distribution of clusters produced by FSI in INCL (solid line) and visible energy loss by ionization (with Birks correction) as a function of their momentum (scatter plot).}
\end{figure}

Figure~\ref{fig:e_step} shows the length of the tracks produced by different particles as a function of their visible energy, as defined in Eq.~\ref{eq:evis}. The sparse population of events at lower track lengths correspond to inelastic processes, which are included in the simulation but not in the characterization of visible energy, as explained in Sec.~\ref{sec:analysis}. In figure~\ref{fig:dEdX_birks} we show the momentum distribution and the deposited  ionization energy  with Birks correction as a function of momentum. 
These distributions are the result of the simulation and the input to the algorithm described in Sec.~\ref{sec:analysis}.
Table~\ref{tab:PIDtr} shows the fraction of events where clusters travel more than 1 or 3~cm in the detector. While $\alpha$ and $^3$He do not travel enough and will mostly contribute to vertex activity (energy deposited around the vertex), deuteron and tritium leave a visible track in a sizable fraction of events and, if searched for, can be quite cleanly identified. 

The primary source of misidentification are secondary interactions through inelastic processes, which typically happen at the end of the cluster track. In such inelastic interactions the cluster looses a large amount of energy and breaks the nucleus. The observed energy released by ionization along the cluster track, before the end of the track, does not include the energy released by secondary interactions and it is therefore  smaller than what is expected for a cluster of that energy. As a consequence, particles could be misidentified with less ionizing ones. Therefore, tritium could be misidentified ($\sim$10\% of events) as proton or deuteron with similar probability, deuteron could be misidentified ($\sim$20\% of events) as protons, while protons cannot be misidentified as nuclear clusters. $^3$He and $\alpha$ rarely travel more than 1 cm and when they do, they can be easily identified due to their very large energy deposit. The characterization of the inelastic interactions, the particles produced and their observability in the detector are left for future studies.

\begin{table}[htbp]
\centering
\caption{\label{tab:PIDtr} Percentage of clusters travelling more than 1 (3) cm} 
\begin{tabular}{|c|c|c|c|c|c|}
\hline
 & \textbf{$\alpha$} & \textbf{$^3$He} & \textbf{T} & \textbf{D} & \textbf{proton} \\
\hline
Travels more than 1 cm, \% & 0.3 & 1.3 & 60 & 72 & 87\\
Travels more than 3 cm, \% & 0   & 0   & 34 & 51 & 74\\
\hline
\hline
\end{tabular}
\end{table}

\begin{figure}[t!]
\centering
\center{\includegraphics[width=0.98\linewidth]{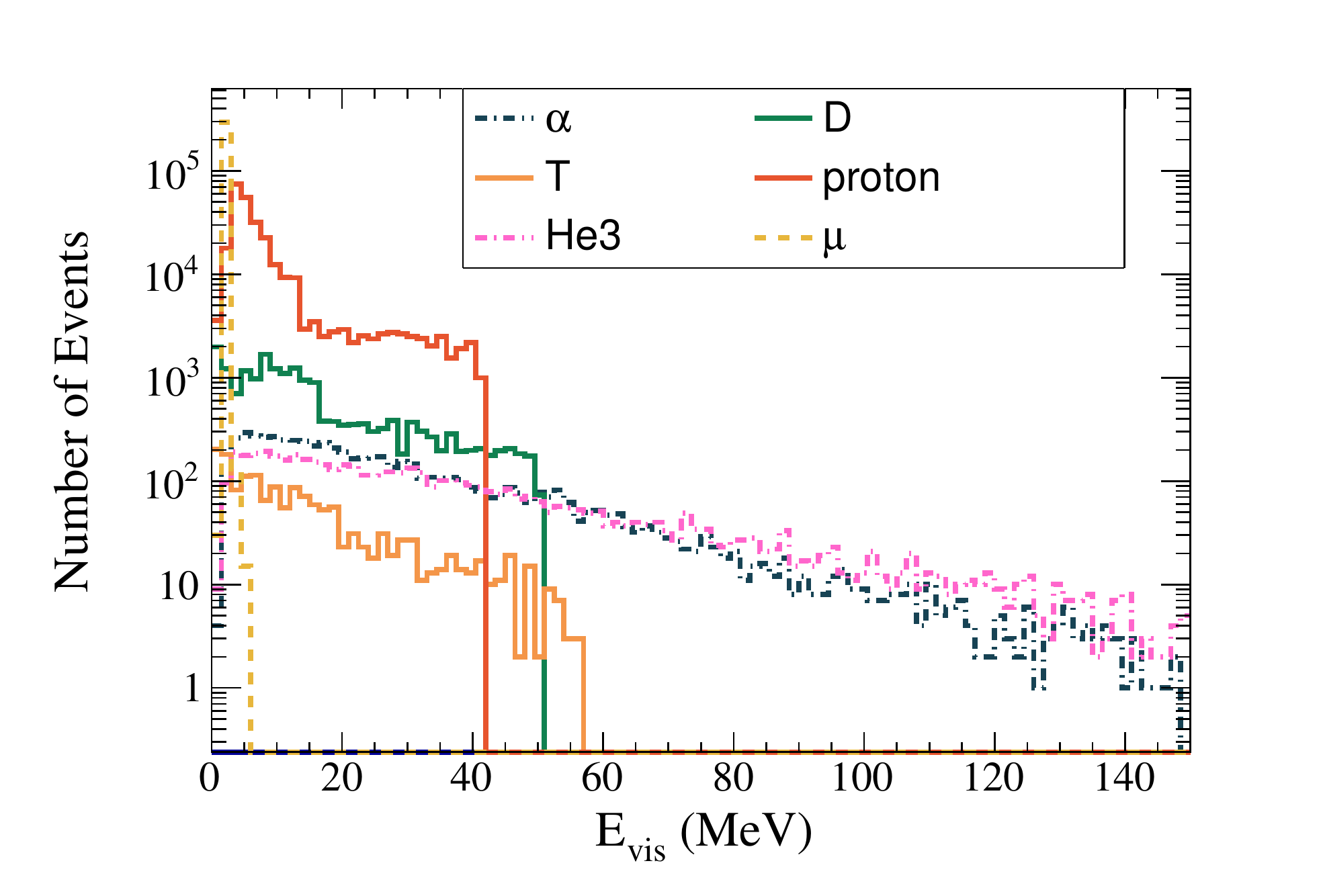} \\}
\caption{\label{fig:VA} Visible energy deposited by ionization by different particles in a sphere of 3~cm radius around the vertex in events simulated with INCL.}
\end{figure}

\begin{figure}[ht]
\centering
\includegraphics[width=0.98\linewidth]{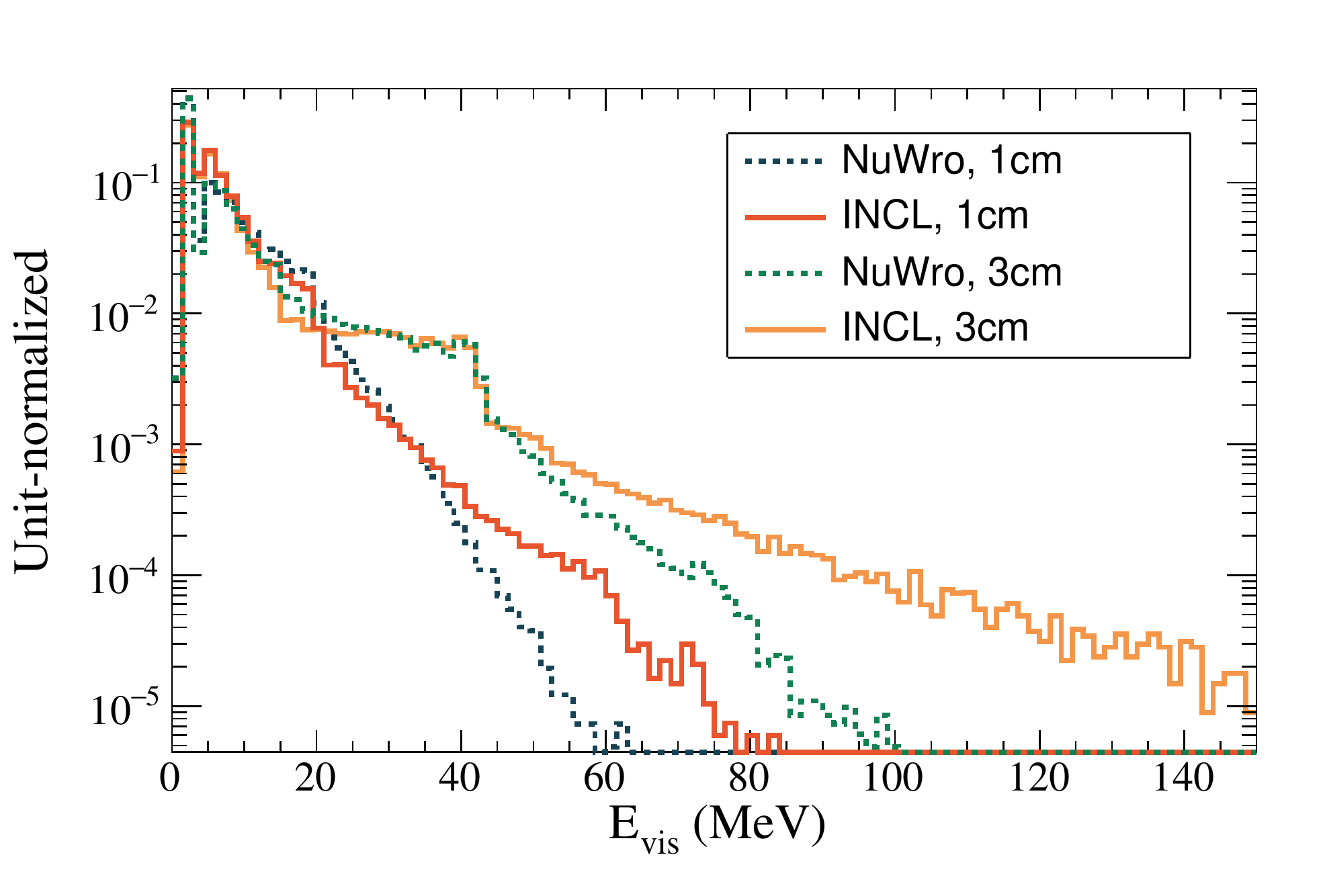}
\caption{\label{fig:VA_all} Total visible energy deposited by ionization by all the particles in a sphere of 1 (3)~cm radius around the vertex. Distributions from events simulated with INCL and NuWro are compared with the same overall normalization.}
\end{figure}

The particles that are below tracking threshold still contribute to the total energy deposited around the vertex. To compute the contribution of clusters to the vertex activity, the visible energy deposited by ionization in a sphere of 1 (3)~cm radius around the vertex is calculated, as shown in Fig.~\ref{fig:VA} and Fig.~\ref{fig:VA_all}. The distributions have two components depending if the particles leave the sphere or if they release all their energy inside the sphere.
The long tail of INCL comes from the energy deposits due to clusters. There are 8~(11)\% of the events with more than 15~(20)~MeV energy deposited in the 1~(3)~cm sphere. The tail of the NuWro distribution comes from multinucleon neutrino interactions and FSI inelastic processes with production of multiple protons. There are 10~(10)\% of the events with more than 15~(20)~MeV energy deposited in the 1~(3)~cm sphere.
The energy deposited by inelastic interactions is not included, since its observability depends on the particles produced by such interactions. Inelastic interactions happen within 1 (3)~cm from the primary vertex in about 3\% (8\%) of tritium and deuteron events, 1\% (3\%) of proton events, while they happen in a negligible fraction of events with $\alpha$ and $^3$He.

\section{Comparison to data}
\label{sec:data}
\begin{figure*}[ht]
\centering
\begin{minipage}[h]{0.49\linewidth}
\center{\includegraphics[width=1\linewidth]{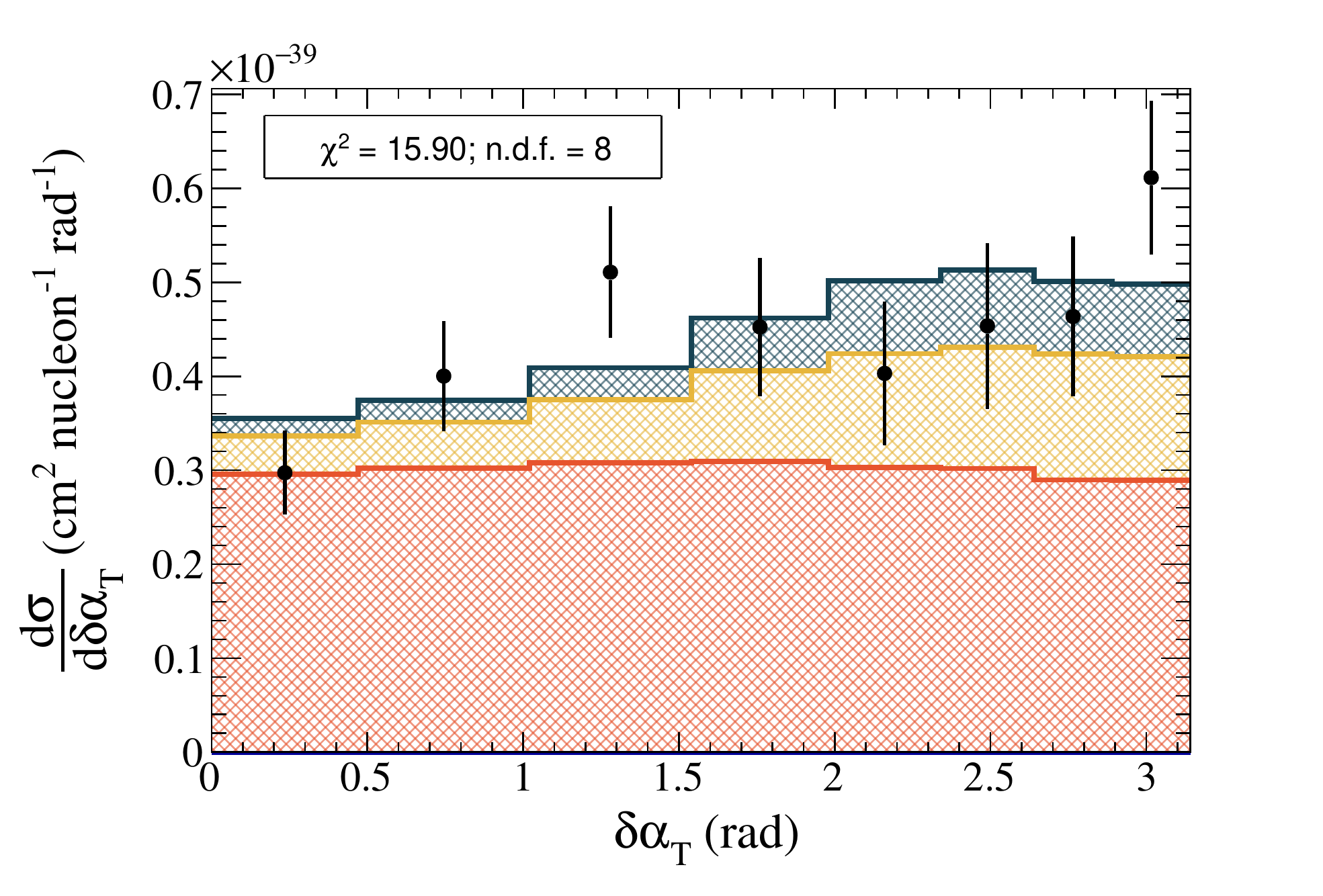} \\}
\end{minipage}
\hfill
\begin{minipage}[h]{0.49\linewidth}
\center{\includegraphics[width=1\linewidth]{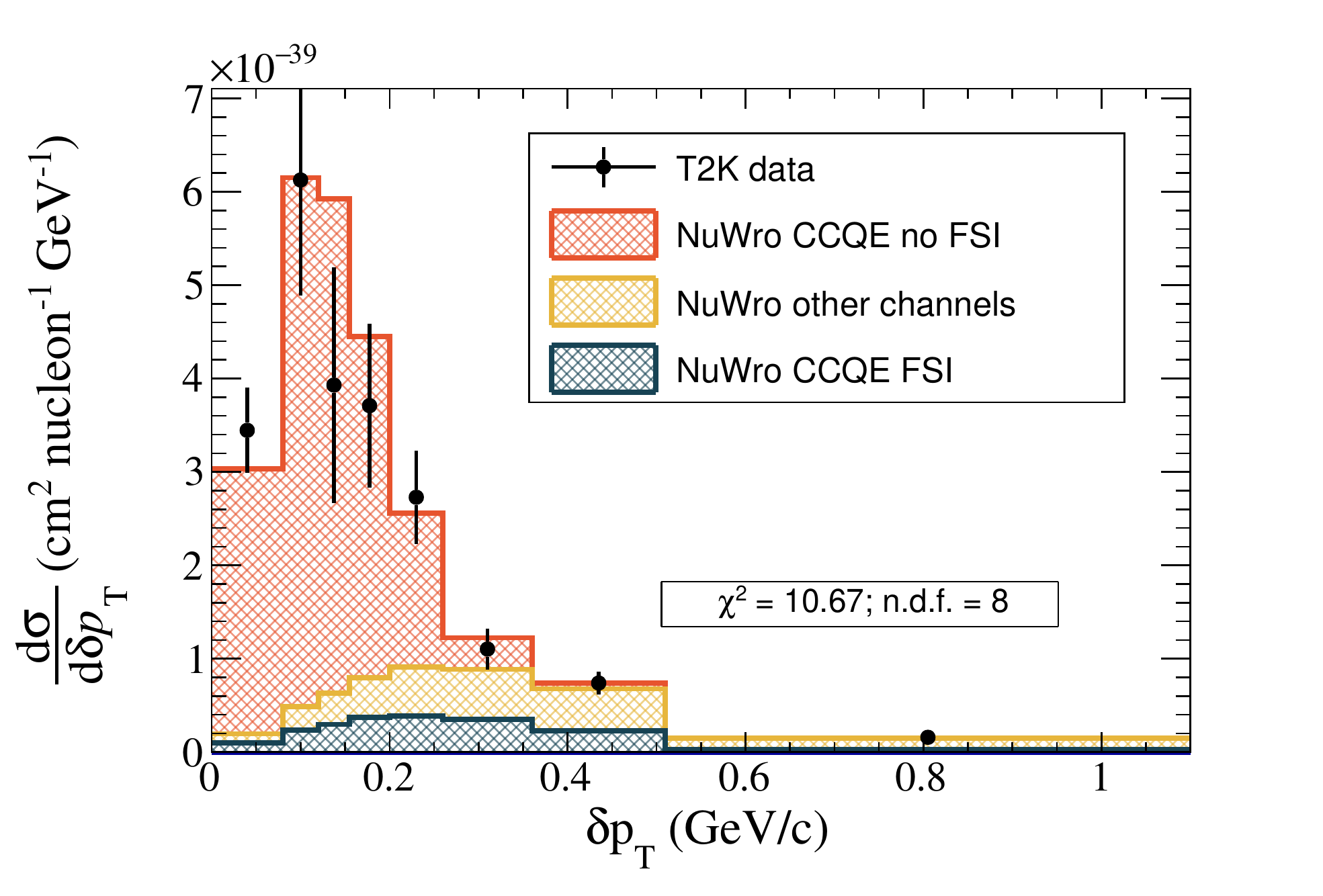} \\}
\end{minipage}
\vfill
\begin{minipage}[h]{0.49\linewidth}
\center{\includegraphics[width=1\linewidth]{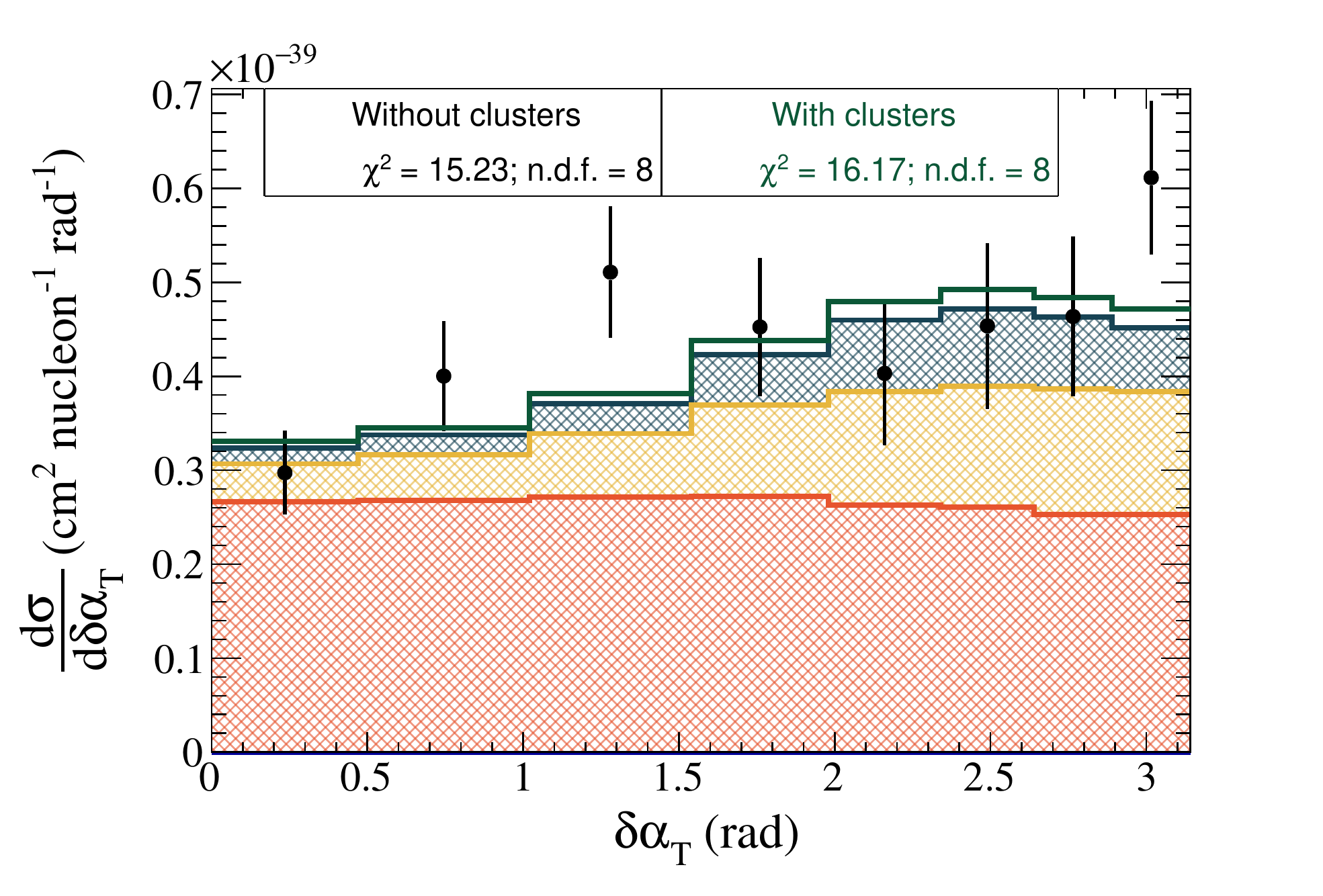} \\}
\end{minipage}
\hfill
\begin{minipage}[h]{0.49\linewidth}
\center{\includegraphics[width=1\linewidth]{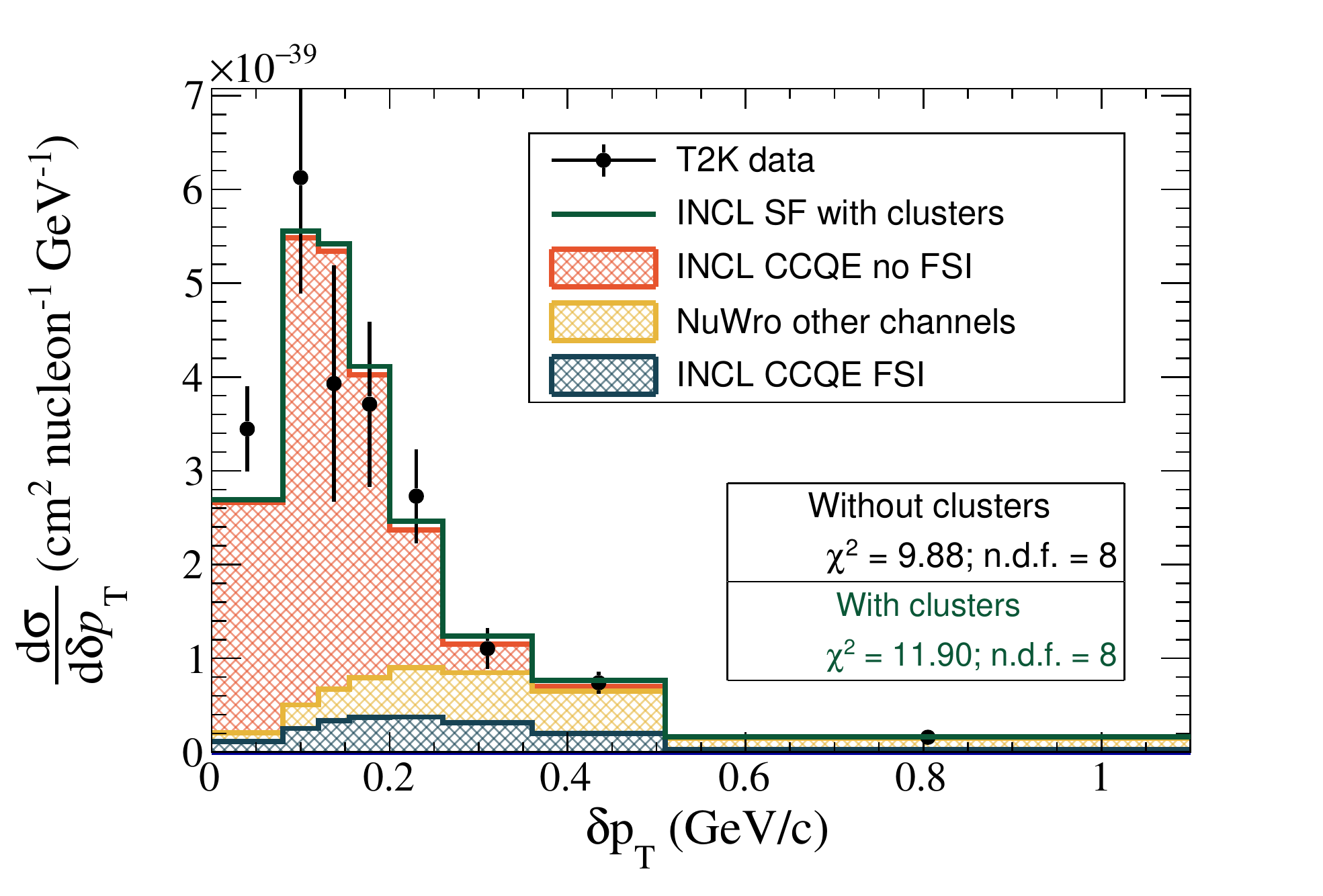} \\}
\end{minipage}
\vfill
\begin{minipage}[h]{0.49\linewidth}
\center{\includegraphics[width=1\linewidth]{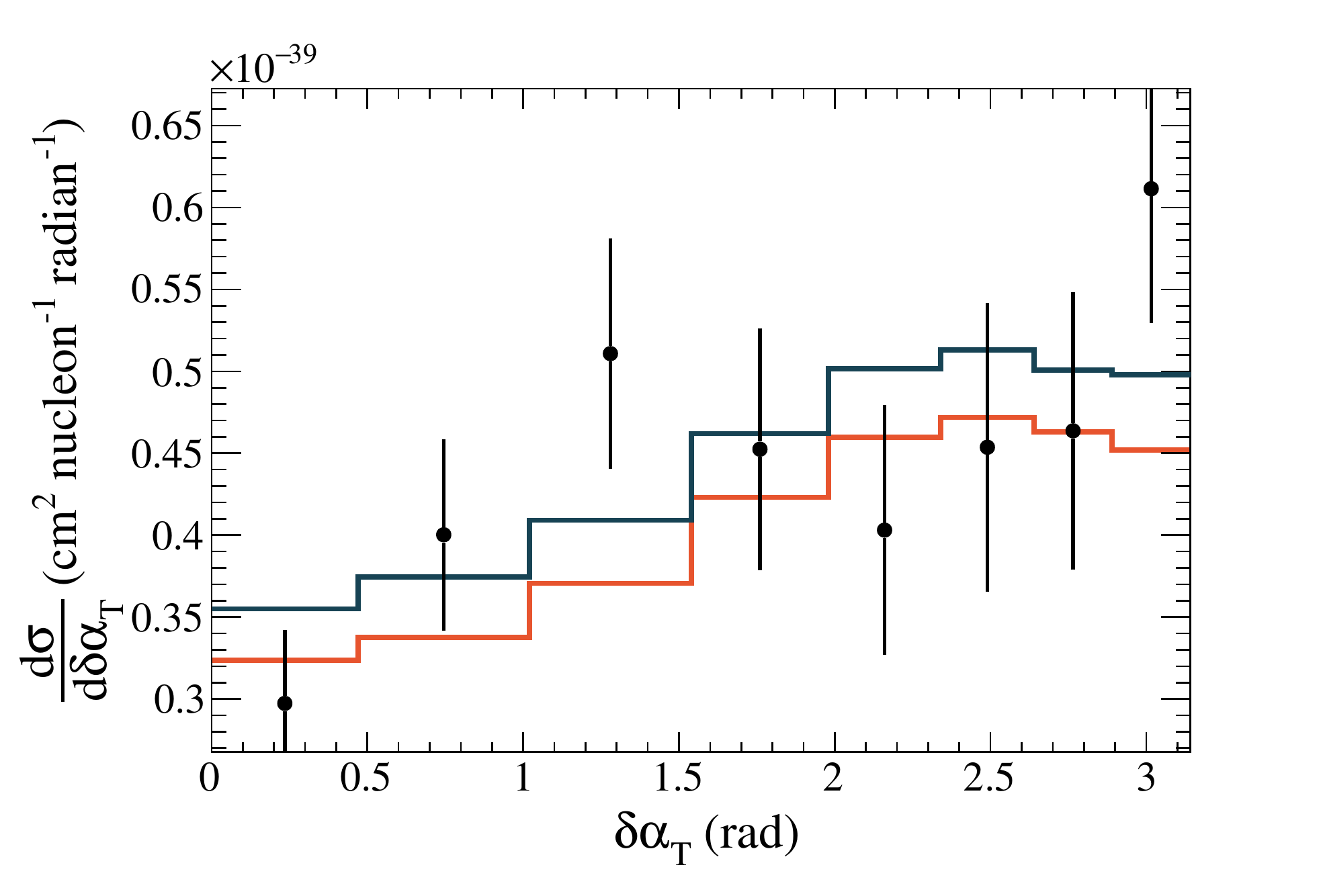} \\}
\end{minipage}
\hfill
\begin{minipage}[h]{0.49\linewidth}
\center{\includegraphics[width=1\linewidth]{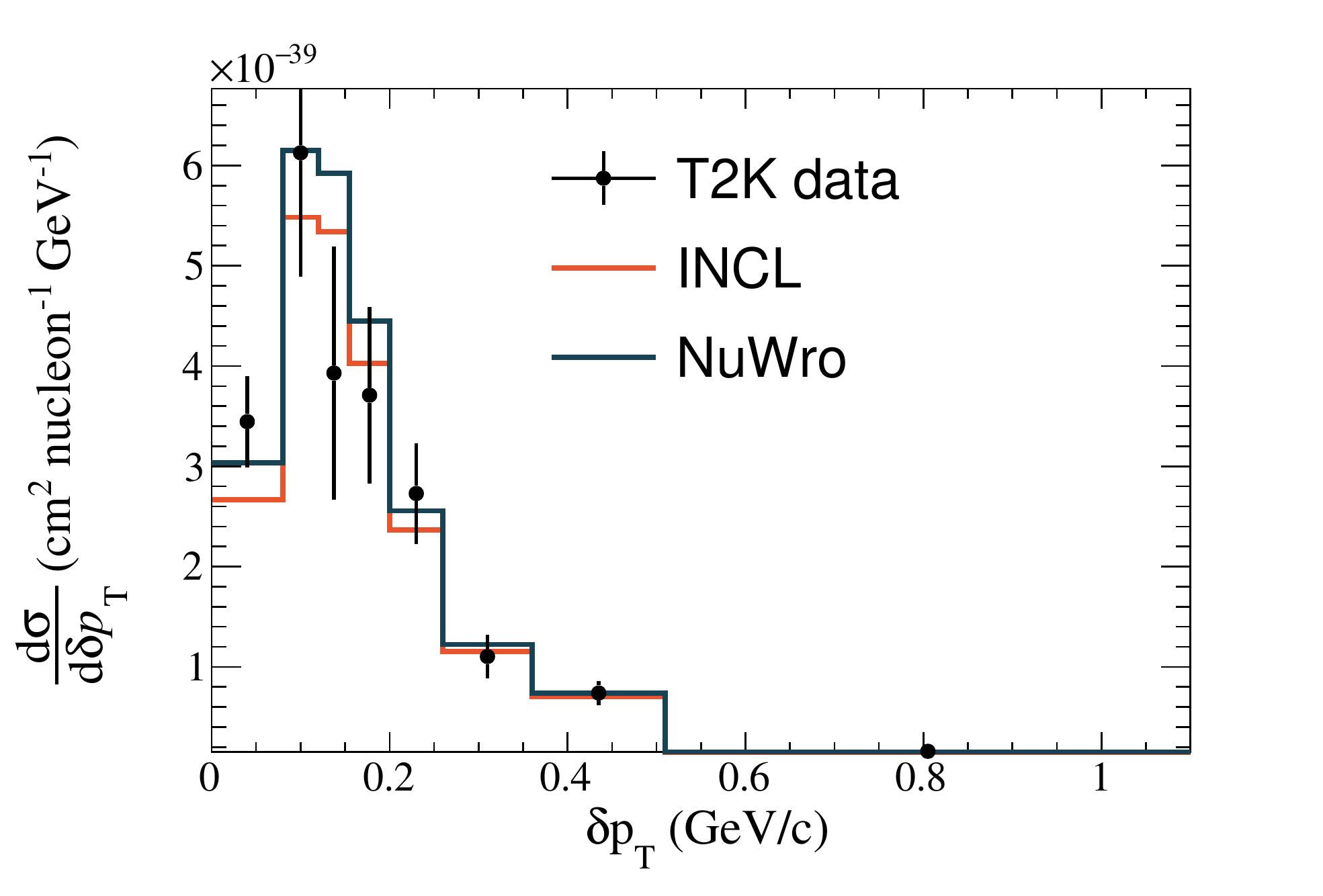} \\}
\end{minipage}
\vfill
\center{\includegraphics[width=1\linewidth]{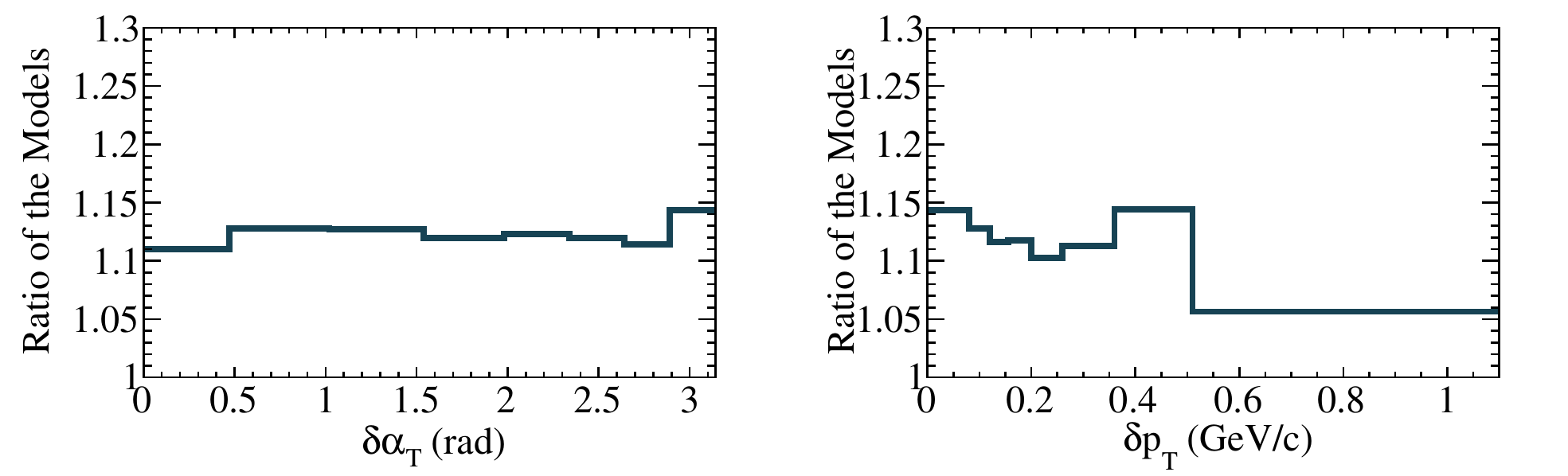} \\}

\caption{\label{fig:dataSF_T2K} Top to bottom:  NuWro SF comparison to T2K data;  INCL + NuWro SF comparison to T2K data; comparison of NuWro, INCL + NuWro SF and data; ratio of NuWro SF and INCL + NuWro SF models of QE channel. Left: $\delta\alpha_T$, right: $\delta$p$_T$}
\end{figure*}

\begin{figure*}[ht]
\centering
\begin{minipage}[h]{0.49\linewidth}
\center{\includegraphics[width=1\linewidth]{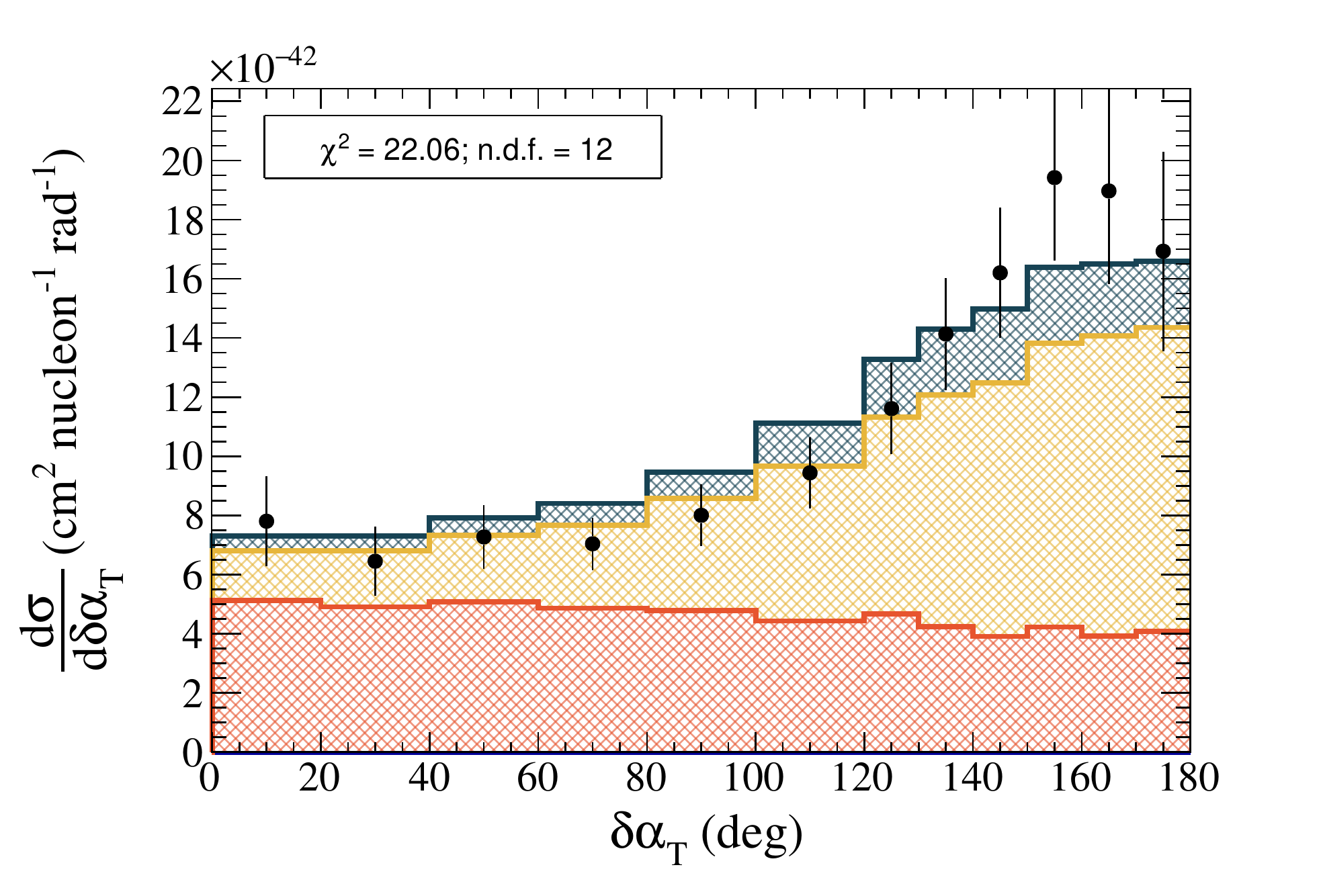} \\}
\end{minipage}
\hfill
\begin{minipage}[h]{0.49\linewidth}
\center{\includegraphics[width=1\linewidth]{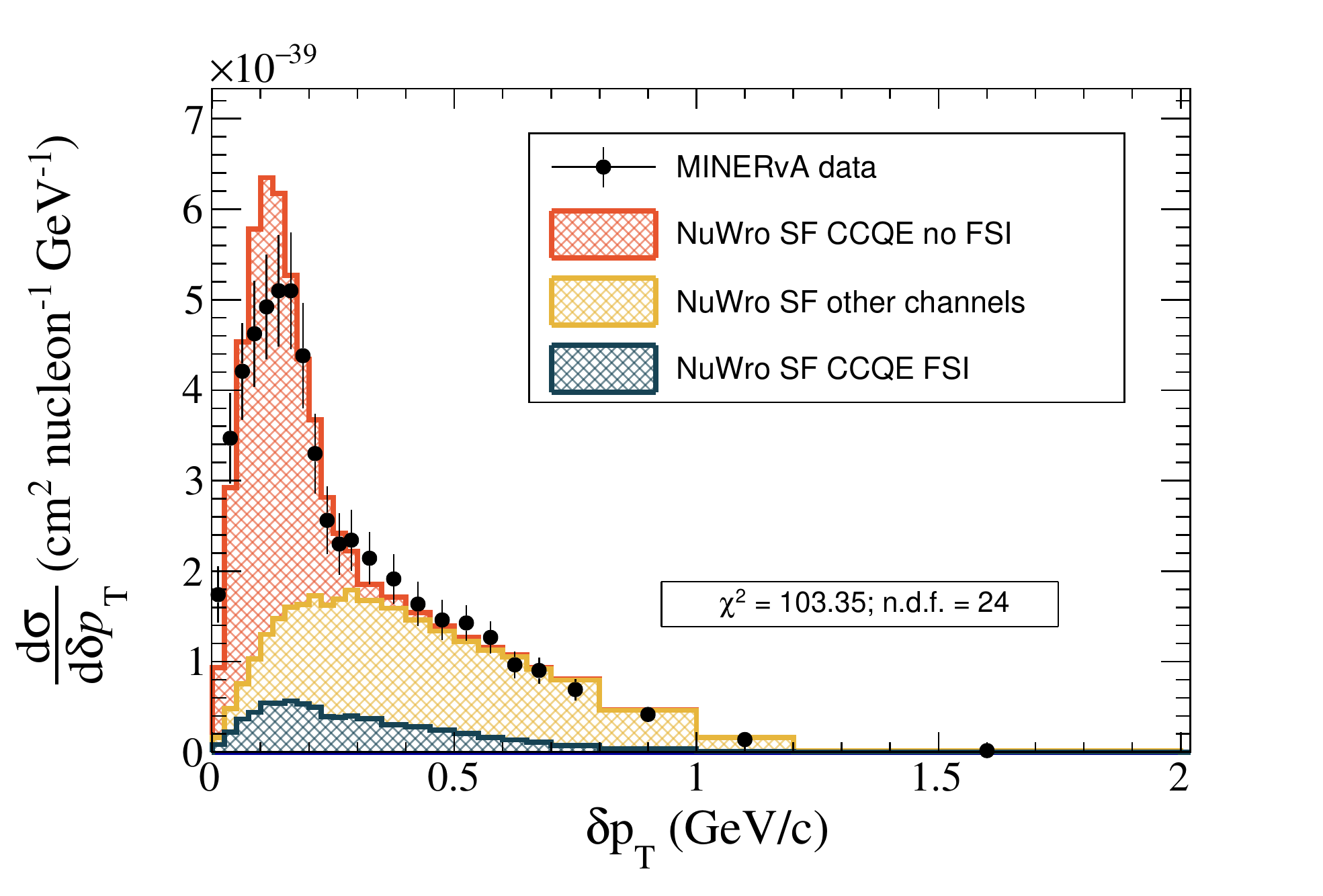} \\}
\end{minipage}
\vfill
\begin{minipage}[h]{0.49\linewidth}
\center{\includegraphics[width=1\linewidth]{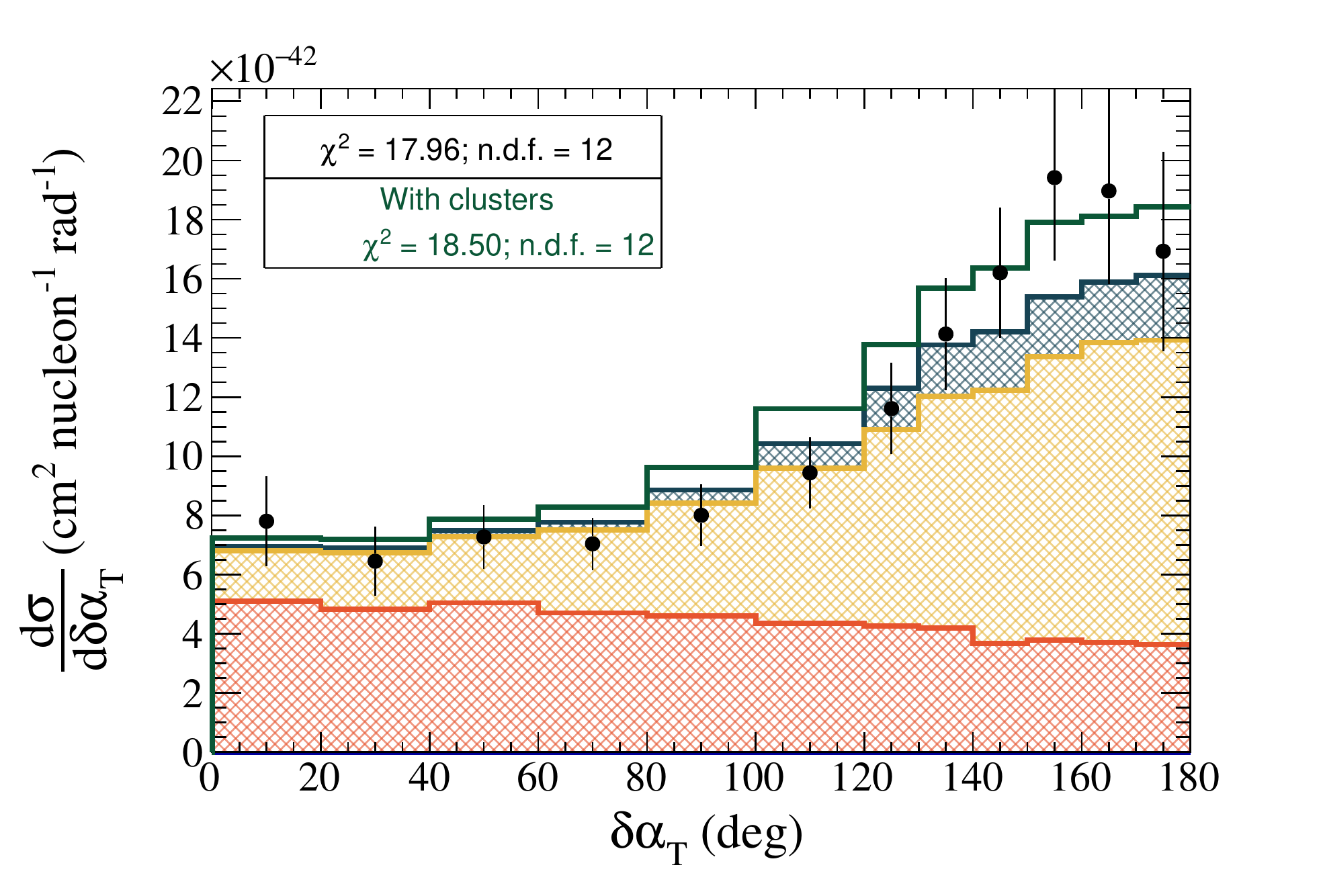} \\}
\end{minipage}
\hfill
\begin{minipage}[h]{0.49\linewidth}
\center{\includegraphics[width=1\linewidth]{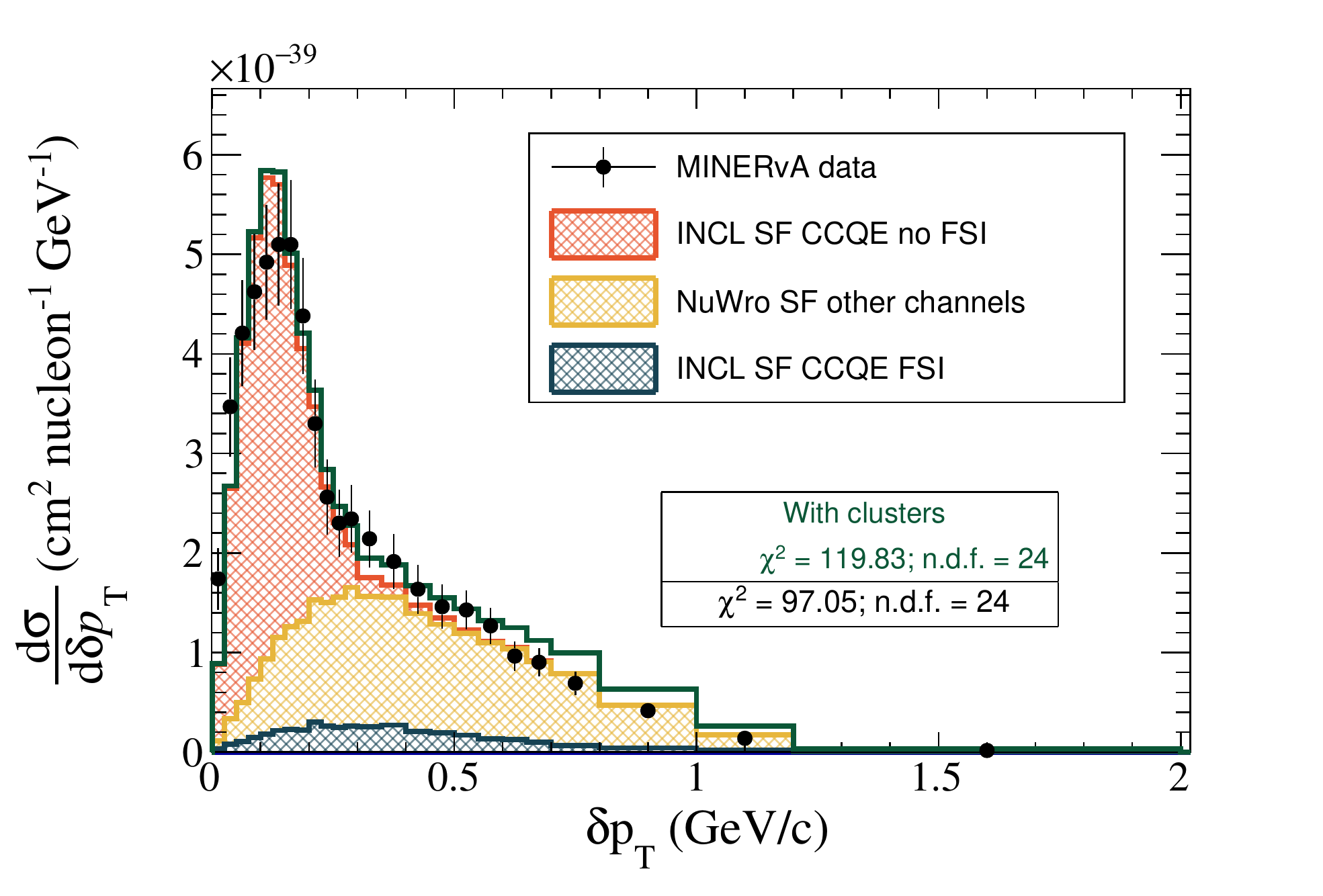} \\}
\end{minipage}
\vfill
\begin{minipage}[h]{0.49\linewidth}
\center{\includegraphics[width=1\linewidth]{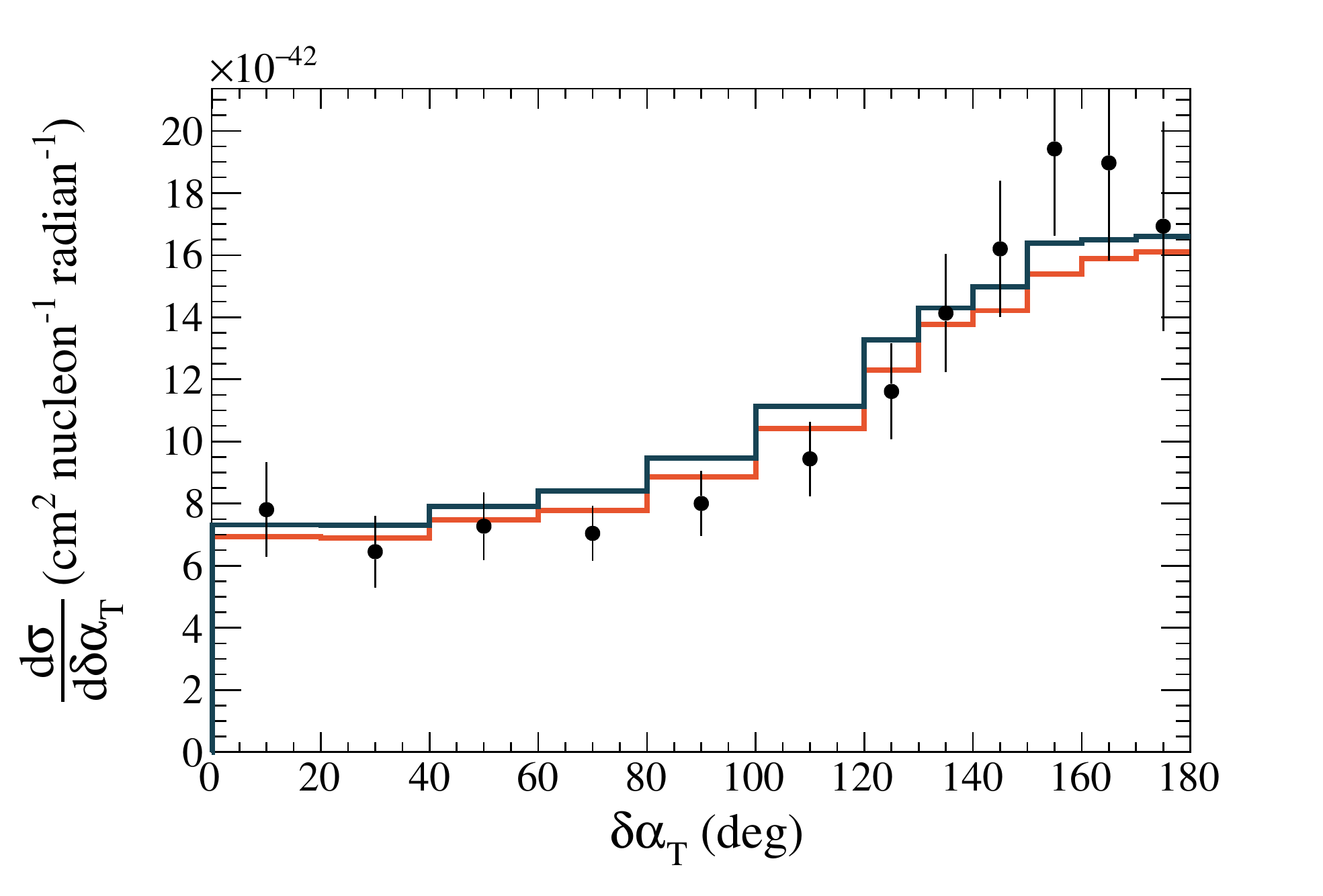} \\}
\end{minipage}
\hfill
\begin{minipage}[h]{0.49\linewidth}
\center{\includegraphics[width=1\linewidth]{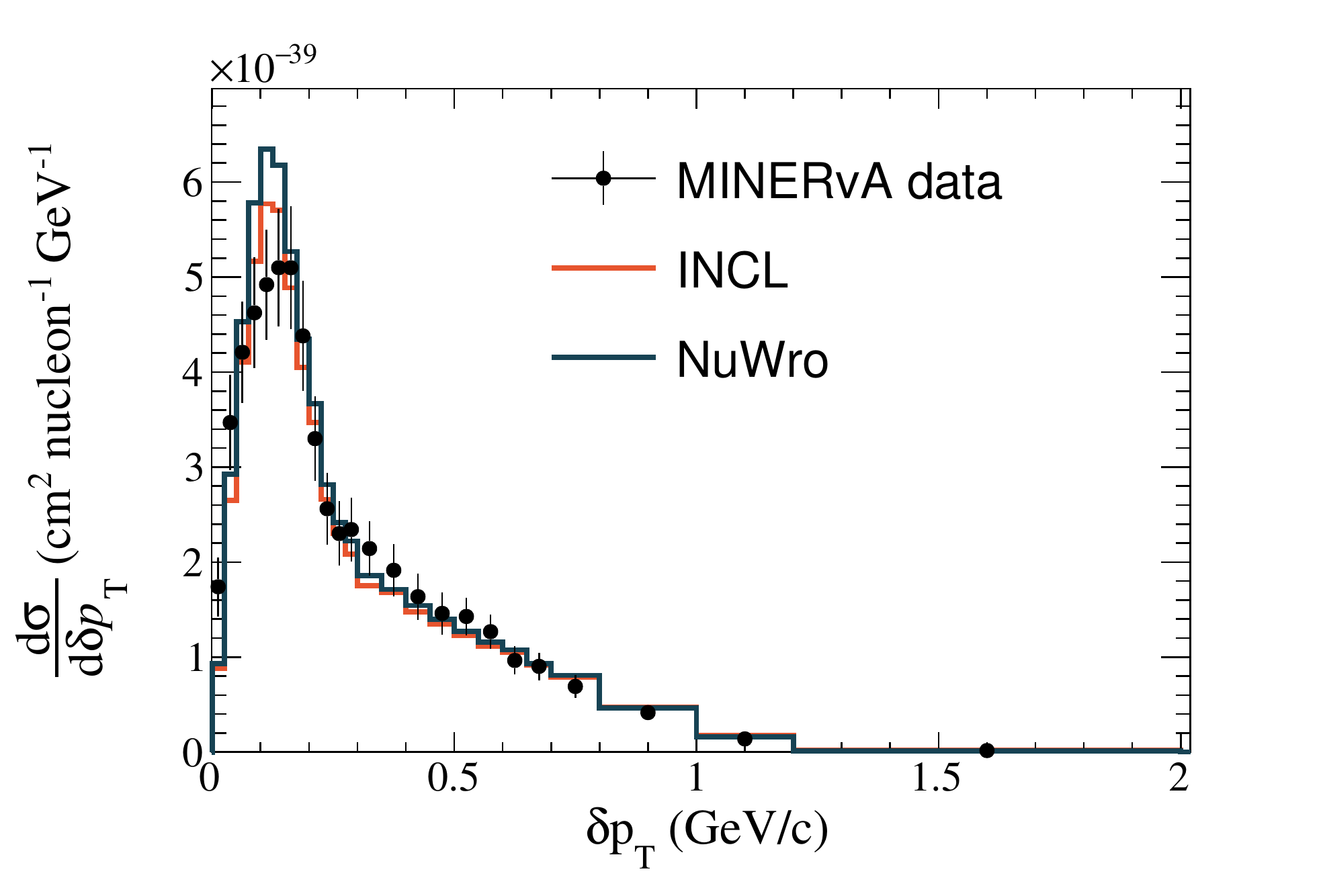} \\}
\end{minipage}
\vfill
\center{\includegraphics[width=1\linewidth]{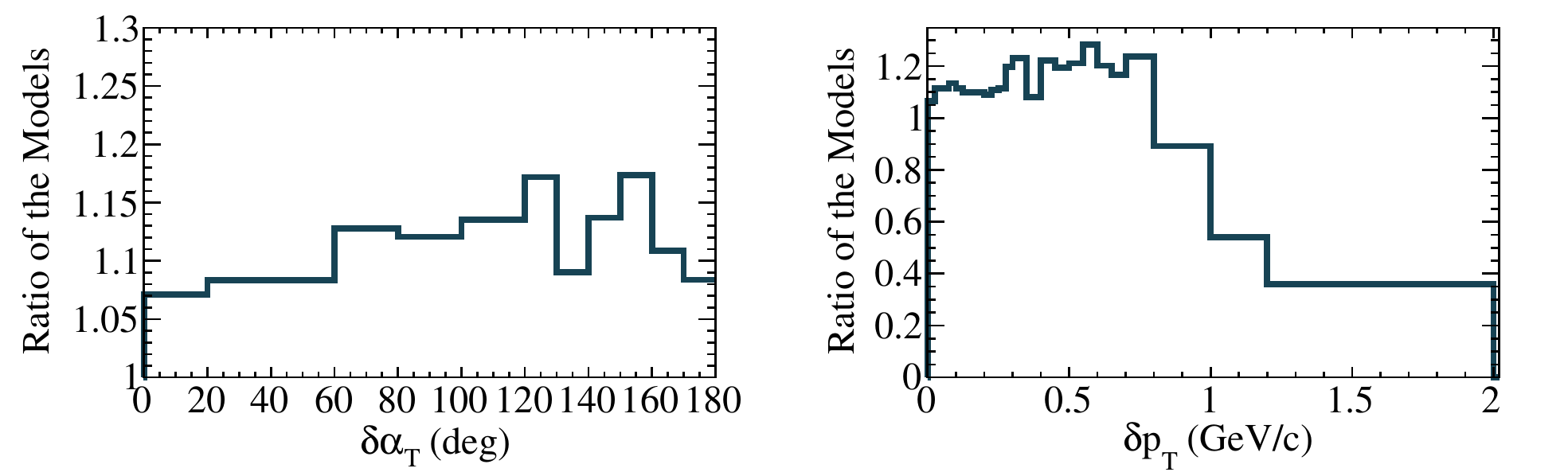} \\}
\caption{\label{fig:dataSF} Top to bottom: NuWro SF comparison to Miner$\nu$a data;  INCL + NuWro SF comparison to Miner$\nu$a data; comparison of NuWro, INCL + NuWro SF and data; ratio of NuWro SF and INCL + NuWro SF models of QE channel. Left: $\delta\alpha_T$, right: $\delta$p$_T$}
\end{figure*}

The prediction of NuWro and INCL are compared to STV measurements ($\delta \alpha_T$ and $\delta p_T$) of T2K~\cite{T2K:2018rnz} and Miner$\nu$a~\cite{MINERvA:2018hba} in Figs.~\ref{fig:dataSF_T2K} and~\ref{fig:dataSF}.
The contribution from non-QE interactions is simulated with NuWro and added on top of INCL predictions. Simulated events are selected to cope with the acceptance of the detectors. Such acceptance cuts, notably requiring large enough proton momentum to be reconstructed, tend to suppress the difference between the models. We look forward to measurements with lower tracking threshold, as expected for instance with ND280 upgrade~\cite{T2K:2019bbb} and the detectors of the SBN program~\cite{Antonello:2015lea}.
The different shape in $\delta\alpha_T$ is washed out, still the lower overall normalization of INCL, due to proton absorption, is visible and tends to slightly improve the comparison to data. The impact of nuclear clusters in the distributions is shown assuming perfect identification and same acceptance as protons. With present momentum threshold, nuclear clusters tend to distribute at high $\delta\alpha_T$, similarly to protons affected by FSI. Given the relatively high momentum threshold for proton tracking, the different proton momentum distributions after FSI in the two cascade models has a subleading effect.

\section{Conclusions}
\label{sec:conclusions}

We have compared the simulation of the final-state interactions between the NuWro and INCL cascade models in CCQE events. We have characterized the produced particles and the change of kinematics of the leading proton induced by FSI. We have quantified the impact on observables like STV and vertex activity. We have also compared the results of the simulations with the existing experimental data provided by T2K and MINER$\nu$A collaborations.

An essential novelty of this study is the simulation of cluster production by INCL in FSI of neutrino interactions. Such clusters can be detected as highly ionizing tracks or, if below the tracking threshold, can contribute to the energy deposited around the vertex (vertex activity). Nuclear clusters behave differently than protons from many points of view. Since nuclear clusters release more energy by ionization, the detector tracking threshold tends to require the clusters to have larger momentum than protons in order to be reconstructed. When identified as tracks, the particle momentum could be measured by its curvature, and the corresponding kinetic energy depends on the actual particle mass. Generally, the kinetic energy of a cluster leaving the nucleus is different from the kinetic energy of the initial proton producing the cluster. 
Moreover, for a complete recollection of all the energy of the particles leaving the nucleus, the correct reconstruction of their secondary interactions is needed. While we leave the detailed characterization of secondary interactions for future work, we have considered the fraction of energy lost in secondary interactions and their probability, which depends on the nature of the particles leaving the nucleus. Notably, emitted nuclear clusters have different probability of secondary interactions with respect to protons.
For all these reasons, the accurate modeling and identification of particles in the final state, including nuclear clusters, are crucial to estimate correctly all the energy leaving the nucleus and the FSI corrections to it.
Thus, the modeling of cluster production by FSI is relevant for correct interpretation of data to constrain nuclear models and it may affect the calorimetric method of total neutrino energy reconstruction and how it is shared between the leptonic and hadronic part of the final state.  The precise neutrino energy reconstruction is a vital step for the new generation of oscillation experiments. The calorimetric method of the neutrino energy reconstruction, which takes into account all emitted particles after the neutrino interaction, allows for better neutrino energy resolution but also demands better modeling of the hadronic compound of the neutrino-nucleus interactions. Understanding the production and kinematical properties of nuclear clusters is crucial for improving the existing neutrino-nucleus interaction models and controlling the corresponding systematic uncertainties.

Regarding the effects of the final-state interactions on the leading proton in neutrino-nucleus interactions, the nuclear model of INCL features much smaller transparency than the NuWro cascade. Transparency is a measure that allows estimating the FSI strength in a given model. Therefore, such results originate in a characteristic combination of the nucleon-nucleon cross sections,  Pauli blocking, and the specifics of the used nuclear model. In particular, INCL FSI simulation features a significant fraction of events without a proton in the final state, especially while propagating low momentum protons from the primary vertex. The INCL model also tends to re-absorb other particles produced during the FSI cascade: events with low momentum protons are mostly accompanied in NuWro with additional nucleons but not in INCL.

The accurate modeling of the rate of events without protons in the final state directly affects the correct interpretation of experimental data in long-baseline neutrino oscillation experiments. Regarding specific observables, the depletion of low momentum protons causes suppression of large values of $\delta \alpha_T$. Unfortunately, the proton tracking threshold in the detector induces a similar effect:  as we show in the comparison to T2K and Minerva measurements, the available data have very limited sensitivity to the difference between the two FSI models due to the proton momentum threshold of present detectors. These results demonstrate the importance of measuring low momentum protons to characterize their FSI. Currently running and upcoming experiments are planning to perform such measurements by providing low tracking thresholds and precise calorimetry~\cite{steven_gardiner_2022_6717687}.

Summarizing, the study presented here is a progress along the road of a complete and precise estimation of uncertainties due to the modeling of final-state interactions effects of neutrino-scattering simulation. It is also a useful example of the modularity of Monte Carlo generators and a step forward toward improving the simulation of FSI in neutrino-nucleus interactions.

\section*{Acknowledgements}
We have conducted this work within the T2K Near Detector upgrade project and gratefully acknowledge all productive meetings with our colleagues in this context.
We appreciate the fruitful discussions at the INCL meetings, particularly those with S.~Leray, D.~Mancusi, D.~Zharenov, and J.~L.~Rodríguez-Sánchez.
This work was supported by P2IO LabEx (ANR-10-LABX-0038 – Project ``BSMNu'') in the framework ``Investissements d’Avenir'' (ANR-11-IDEX-0003-01), managed by the Agence Nationale de la Recherche (ANR), France.
We acknowledge the support of CNRS/IN2P3 and CEA. J.T.~Sobczyk and K.~Niewczas were partially supported by the Polish Ministry of Science and Higher Education grant No. DIR/WK/2017/05, and under the Polish National Science Centre (NCN) grants No. 2021/41/B/ST2/02778 and No. 2020/37/N/ST2/01751, respectively.

\clearpage

\appendix
\section{Initial state nuclear model}
\label{app:pn}
The INCL FSI study presented in this paper relies on an approximated simulation where the initial state is taken from NuWro SF. As discussed in Sec.~\ref{sec:sim}, biases can be induced in the procedure of merging NuWro SF with the INCL nuclear model. The study is repeated using relativistic Fermi gas (RFG) in NuWro to model the initial state, in order to demonstrate that the conclusions on the characterization of FSI in INCL and the differences between NuWro and INCL FSI are robust against assumptions on the initial nuclear state.  
It is well known that RFG model is unable to reproduce the measured distributions of proton kinematics and STV~\cite{T2K:2018rnz,MINERvA:2018hba}. RFG is here used only as a diagnostic tool to test the robustness of the FSI results. In this case, the momentum and position of the nucleon is provided by NuWro RFG and the INCL nuclear state is built around that nucleon. 

\begin{figure}[ht]
\centering
\hfill
\begin{minipage}[h]{1\linewidth}
\center{\includegraphics[width=0.95\linewidth]{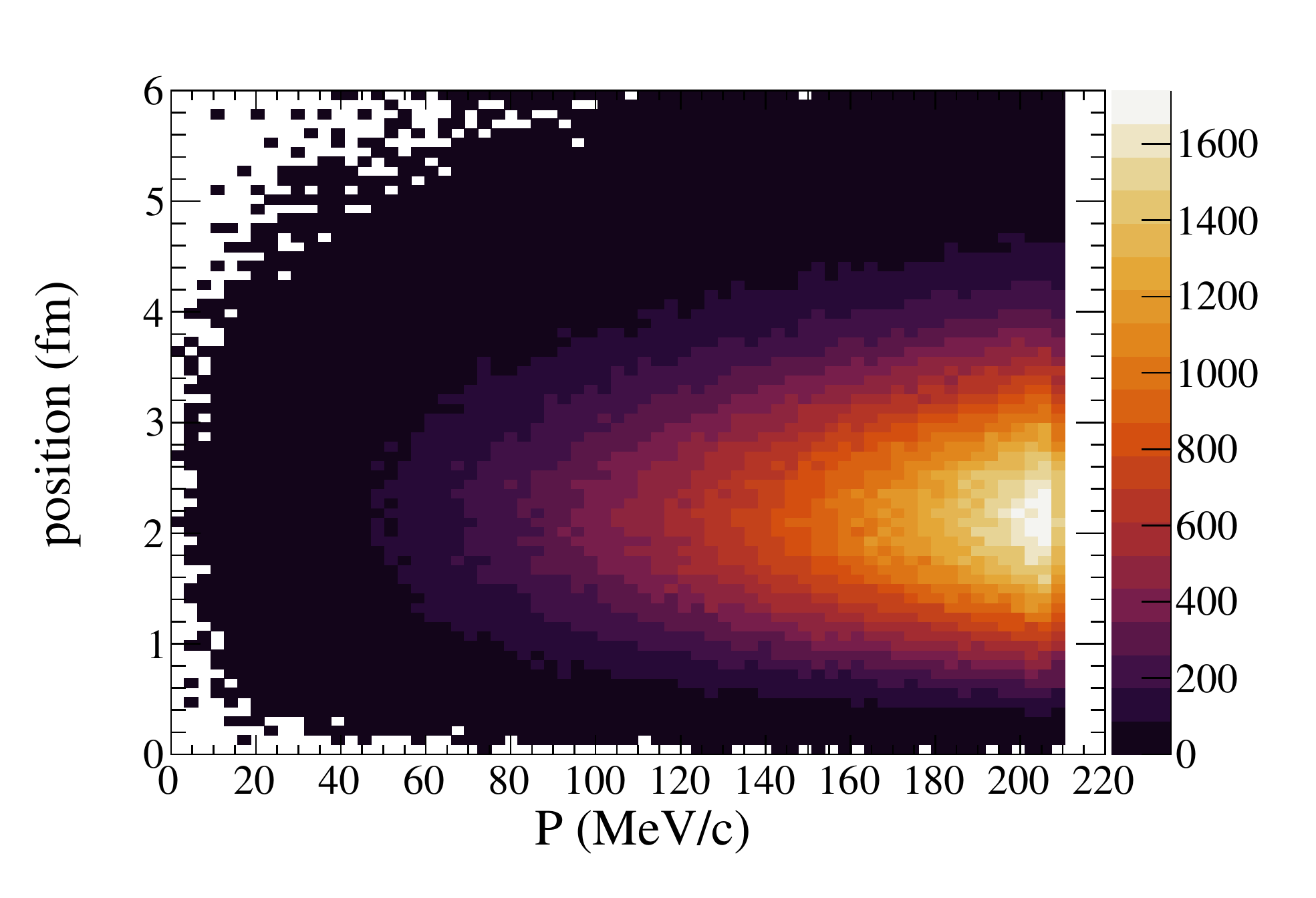} \\}
\end{minipage}
\caption{\label{fig:mom_pos}  Radial coordinate and momentum distribution inside the nucleus for NuWro RFG (z axis in arbitrary units).}
\end{figure}

We are also interested to test a more coherent simulation where both the initial state and the final-state nuclear effects are simulated with the INCL model. To this aim, the NuWro RFG simulation is reweighted to represent the INCL distribution of nucleons momenta and position shown in Fig.~\ref{fig:mom_pos}. The binding energy (for each nucleon position and momentum) and the total cross section are kept as in the original NuWro RFG model.

In Tab.~\ref{tab:channelsRFG}, the fractions of different FSI channels are similar to those observed with SF simulation in Tab.~\ref{tab:channelsSF}.
The fractions change slightly due to the different momentum distribution of proton before FSI but similar trends are observable.

\begin{figure}[ht]
\centering
\center{\includegraphics[width=1\linewidth]{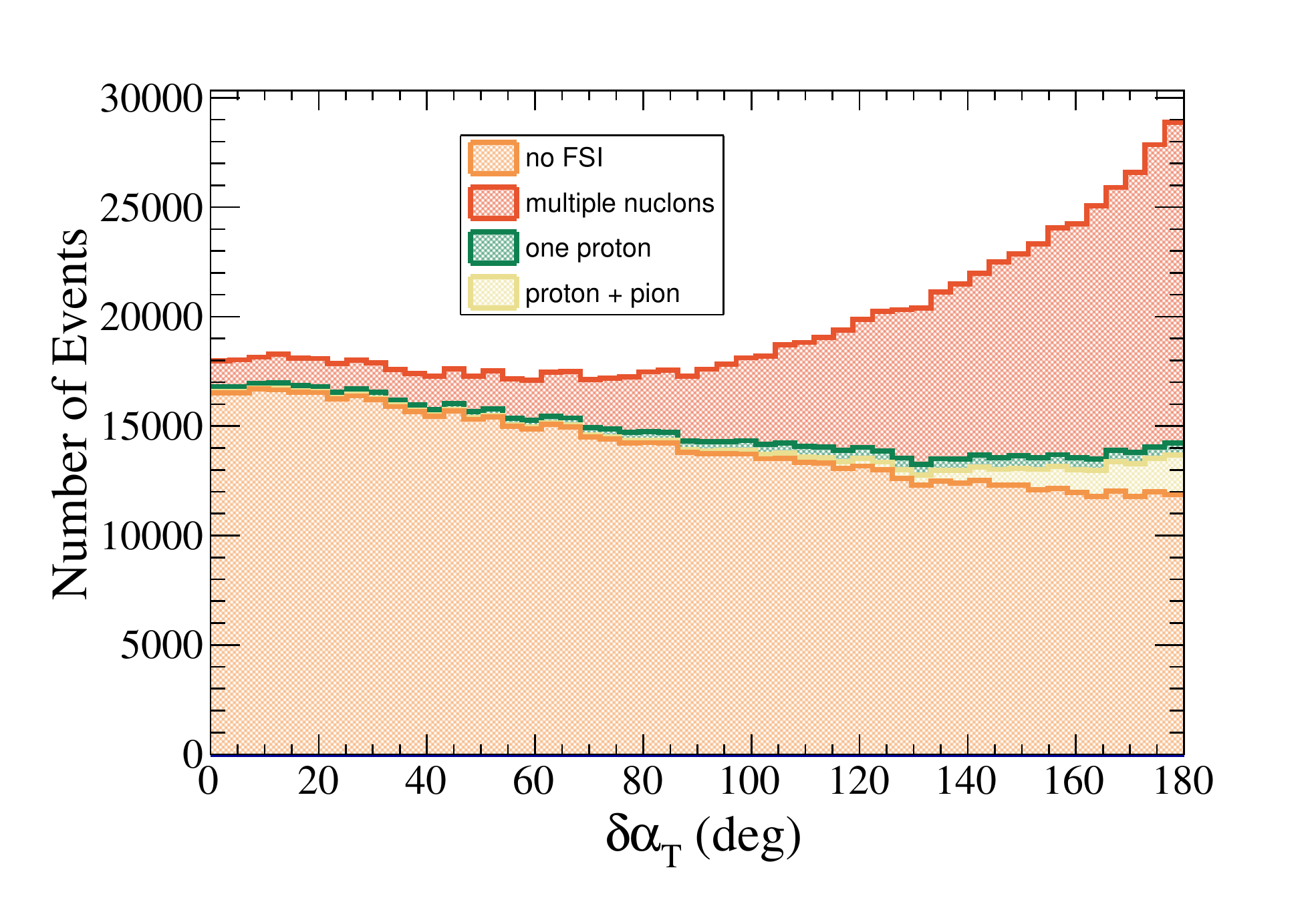} \\}
\center{\includegraphics[width=1\linewidth]{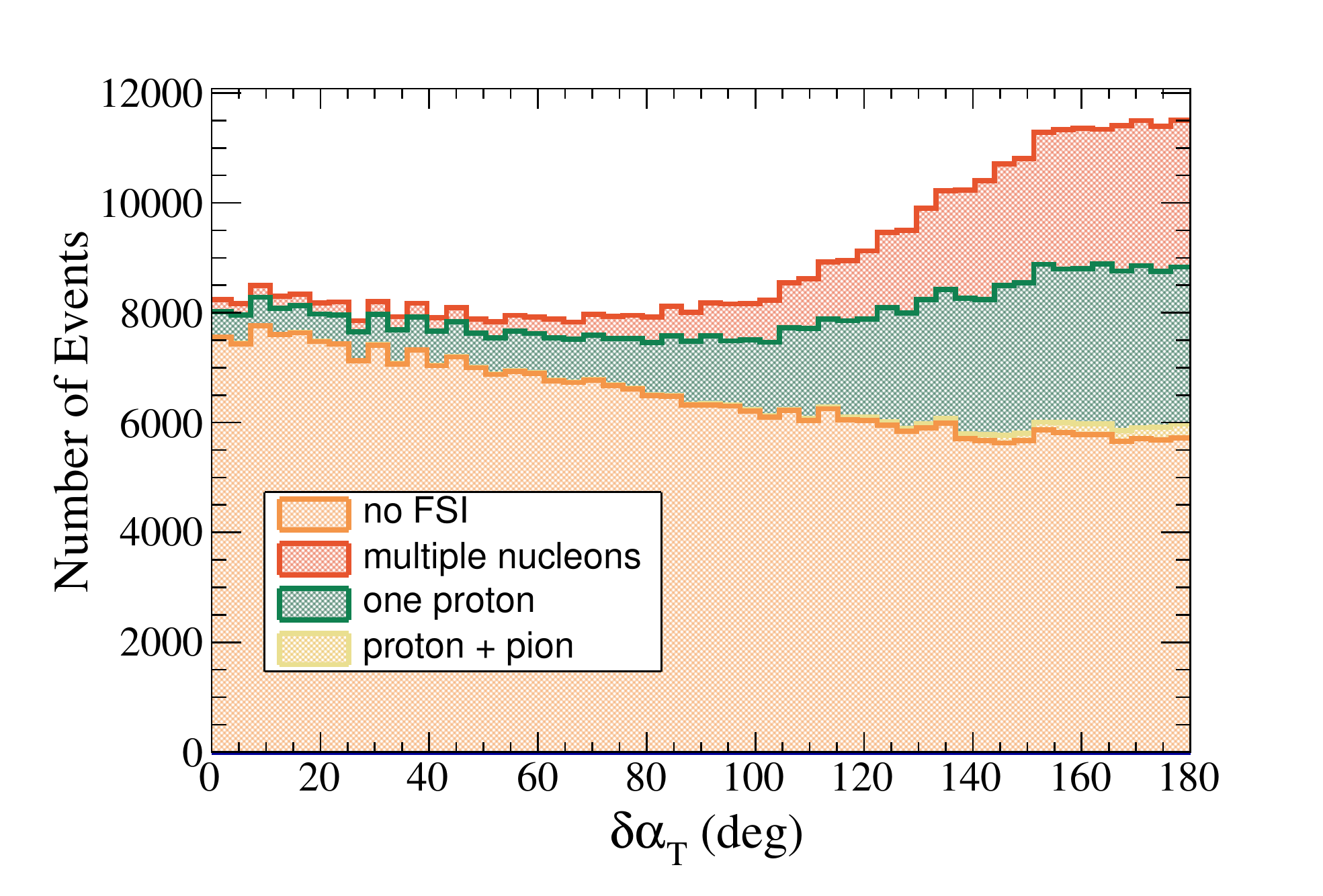} \\}
\center{\includegraphics[width=1\linewidth]{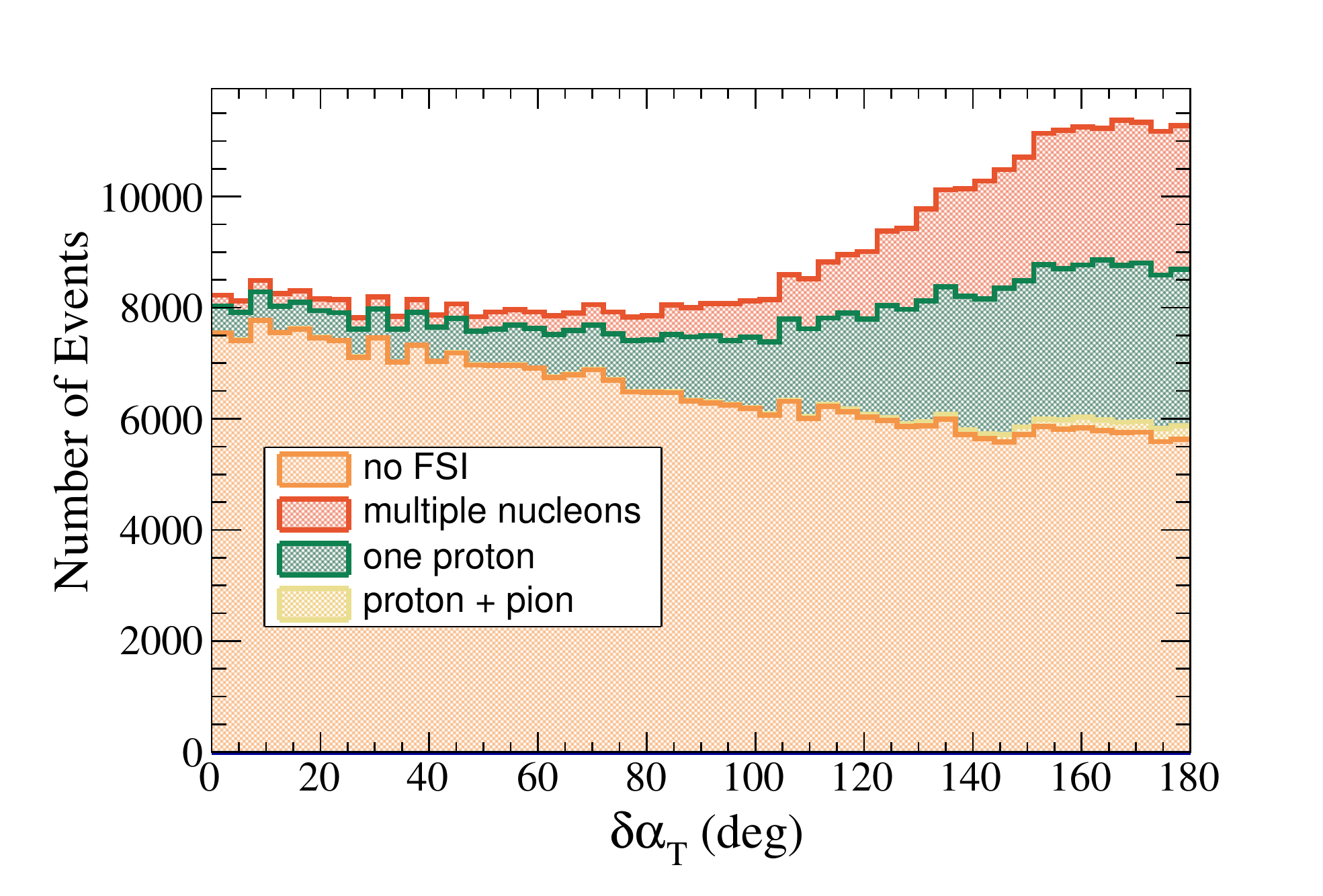} \\}
\caption{\label{fig:rfg} $\delta\alpha_T$ for NuWro RFG (left), INCL+NuWro RFG (middle) and INCL+NuWro RFG reweighted to INCL intial momentum and position (right).}
\end{figure}

In Fig~\ref{fig:rfg}, $\delta\alpha_T$ distributions are shown. RFG features an enhancement of small $\delta\alpha_T$ values due to the large fraction of protons with large momentum, just below the Fermi limits. Still the characteristic suppression of such region induced by FSI in INCL is visible.

\clearpage

\begin{table*}[t!]
\centering
\caption{\label{tab:channelsRFG} Fractions of events with and without protons in the final state, fractions of different channels in events with and without protons in the final state.}
\begin{tabular}{|c|c|c|c|c|}
\hline
&\textbf{Channel} & \textbf{NuWro RFG} & \textbf{INCL+NuWro RFG} & \textbf{INCL+NuWro reweighted}\\\hline
& no protons  &0.92\% &14.90\% & 14.66\% \\
& protons &  99.08\%  &85.10\% & 85.34\% \\
\hline
\hline
\multicolumn{1}{|c|}{\multirow{4}{*}{\rotatebox{90}{\textbf{0 proton}}}}& absorption   & 1.12\% & 25.84\% & 25.89\% \\ 
\multicolumn{1}{|c|}{}&neutron + $\pi$ production  & 4.63\%  & 0.85\% & 0.86\% \\  
\multicolumn{1}{|c|}{}&$\pi$ production& 0.04\%              &    0\% & 0\%\\

\multicolumn{1}{|c|}{}&neutron production  & 94.21\% & 36.655\% & 36.55\%\\
\multicolumn{1}{|c|}{}&cluster production & 0\%      & 36.655\% & 36.70\% \\
\hline
\hline
\multicolumn{1}{|c|}{\multirow{4}{*}{\rotatebox{90}{\textbf{proton}}}}& no FSI  & 71.47\% & 72.03\% & 72.53\%\\ 
\multicolumn{1}{|c|}{}&1 proton         &  2.0\%   & 16.08\% & 15.8\% \\  
\multicolumn{1}{|c|}{}&multiple nucleons&  24.63\% & 11.27\% & 11.1\%\\
\multicolumn{1}{|c|}{}&$\pi$ production &  1.9\%   &  0.62\% & 0.57\% \\
\hline
\hline
\end{tabular}
\end{table*}

\bibliography{biblio}

\end{document}